\documentclass{article}
\usepackage{amssymb,amsmath}
\numberwithin{equation}{section}
\newtheorem{lemma}{Lemma}[section]
\newtheorem{definition}[lemma]{Definition}
\newtheorem{proposition}[lemma]{Proposition}
\newtheorem{theorem}[lemma]{Theorem}
\newtheorem{remark}[lemma]{Remark}

\def\supp{{\rm supp}}

\begin{document}
\title{Local existence and continuation criteria for solutions of the
Einstein-Vlasov-scalar field system with surface symmetry}
\author{D. Tegankong$^{1}$; N. Noutchegueme$^{2}$;
A.D. Rendall$^{3}$\\
$^{1}$Department of Mathematics, Higher Teachers Training
College,\\
University of Yaounde 1, Box 47, Yaounde, Cameroon \\
{tegankon@uycdc.uninet.cm}\\
 $^{2}$Department of Mathematics, Faculty of
Sciences,\\ University of Yaounde 1, Box 812, Yaounde, Cameroon \\
{nnoutch@uycdc.uninet.cm} \\
$^{3}$Max-Planck-Institut f\"ur
Gravitationsphysik,\\ Am M\"uhlenberg 1, D-14476 Golm, Germany \\
{rendall@aei.mpg.de} }
\date{}
\maketitle
\begin{abstract}
We prove in the cases of spherical, plane and hyperbolic symmetry
a local in time existence theorem and continuation criteria for
cosmological solutions of the Einstein-Vlasov-scalar field system,
with the sources generated by a distribution function and a scalar
field, subject to the Vlasov and wave equations respectively. This
system describes the evolution of self-gravitating collisionless
matter and scalar waves within the context of general relativity.
In the case where the only source is a scalar field it is shown
that a global existence result can be deduced from the general
theorem.
\end{abstract}
\section{Introduction}
In the mathematical study of general relativity, one of the main
problems is to establish the existence and properties of global
solutions of the Einstein equations coupled to various matter
fields such as collisionless matter described by the Vlasov
equation (see \cite{andreasson02a}, \cite{rendall02a} for reviews)
or a scalar field (see \cite{rendall95} for the cosmological case
and \cite{christodoulou93} and references therein for the
asymptotically flat case). An essential tool in an investigation
of this type is a local in time existence theorem together with a
continuation criterion. In this paper a theorem of this kind is
proved for one particular choice of matter model.

In \cite{rein1}  and \cite{rein} G. Rein obtained cosmological
solutions of the Einstein-Vlasov system with surface symmetry
written in areal coordinates. In \cite{tchapnda1} and
\cite{tchapnda}, these results were generalized to the case of
non-vanishing cosmological constant. In the present paper, we
extend the results of \cite{rein1} on local existence and
continuation criteria to the case where the source terms of the
Einstein equations are generated by both a distribution function
$f$ of particles, which is subject to the Vlasov equation, and a
massless scalar field $\phi$, which is subject to the wave
equation.

There are several reasons why it is of interest to look at the
case of a scalar field. The first is that it is the simplest
situation in which wave phenomena can be examined in the context
of the Einstein-Vlasov system. In surface symmetry all wave
propagation can be eliminated from the Einstein equations by the
use of suitable coordinate conditions. This is an analogue of the
well-known statement that there are no gravitational waves in
spherical symmetry. Mathematically it means that controlling
solutions of the Einstein equations can be reduced to controlling
solutions of ordinary differential equations in time and in space.
The Vlasov equation, being a scalar hyperbolic equation of first
order, can also easily be solved in terms of its characteristics.
This was the strategy used in \cite{rein1} and \cite{rein}. In the
presence of a cosmological constant it is possible to follow the
same route.

The inclusion of a scalar field introduces waves into the system
which cannot be eliminated. Mathematically this means that it
introduces a non-trivial hyperbolic equation, the wave equation.
This paper is concerned with symmetric situations where there is a
symmetry group acting on two-dimensional spacelike orbits. This
means that the wave equation reduces to an effective equation in
one space dimension. As a consequence part of the strategy used
previously can be carried over. That was based on pointwise
estimates and not on integral estimates (energy estimates) as is
usual in the theory of hyperbolic equations. Pointwise estimates
for solutions of wave equations in terms of data can be obtained
in one space dimension but not in higher space dimensions (See
e.g. the discussion in \cite{kichenassamy}, p. 14.)

Pointwise estimates for hyperbolic equations in one space
dimension can be obtained using the method of characteristics.
This means that in fact ordinary differential equations appear
once again but this time they are integrated not at a constant
value of the spatial or time variable but along characteristics.
This method will be applied in the following, the characteristics
in this case being null curves of the spacetime geometry.

It should be mentioned that there are global existence results in
the literature where the Einstein-Vlasov system is considered in a
context where hyperbolic equations play an important role. In fact
if we relax the assumptions from surface symmetry, where there are
three local Killing vectors, to the case where there are only two
local spacelike Killing vectors, then hyperbolic equations
necessarily occur. Some relevant papers are \cite{rendall97},
\cite{andreasson02b}, \cite{weaver03}. In those references no
direct local existence proof was given. Instead an indirect
argument was used. First a known local existence theorem for the
Einstein-Vlasov equation without symmetry was quoted. Then it was
shown that coordinates could be introduced which are well-adapted
to making use of the symmetry when proceeding to obtain global
results. Apart from the methodological interest of having a direct
local existence proof, the direct proof gives stronger results
concerning the differentiability required of the initial data and
obtained for the solutions. This is very difficult to control in
the indirect method and for this reason the latter has only been
applied to the case where everything is of infinite
differentiability.

The inclusion of a scalar field can be seen as a step towards
certain questions of physical interest. In recent years
cosmological models with accelerated expansion have become a very
active research topic in response to new astronomical observations
\cite{straumann}. The easiest way to obtain models with
accelerated expansion is to introduce a positive cosmological
constant, a possibility studied mathematically in
\cite{tchapnda1}, \cite{tchapnda} and \cite{lee}. A more
sophisticated way is to introduce a scalar field with potential
(see \cite{rendall02b}, section 4.3., \cite{rendall04}, \cite{lee04}).
In the present paper only a
linear scalar field is treated but it is likely that the approach
developed here will be useful in the nonlinear case. Scalar fields
also play a role in theories of gravity generalizing Einstein's
theory, such as the Jordan-Brans-Dicke theory. In that case, in
contrast to the one considered here, there is a direct coupling
between the scalar field and the distribution function. The
techniques developed here could serve as a first step towards the
study of these more complicated situations, which have hardly been
looked at mathematically yet. (See however \cite{andreasson03} and
\cite{calogero03} where the coupling of the Vlasov equation to a
scalar field of the Jordan-Brans-Dicke type in the absence of
Einstein gravity is considered.)

 We consider a four-dimensional spacetime manifold $M$, with local
coordinates $(x^{\alpha})= (t,x^{i})$ on which $x^{0}=t$ denotes
the time and $(x^{i})$ the space coordinates. Greek indices always
run from $0$ to $3$, and Latin ones from $1$ to $3$. On $M$, a
Lorentzian metric $g$ is given with signature $(-,+,+,+)$.
 We consider a self-gravitating collisionless gas and restrict ourselves to the
case where all particles have the same rest mass, normalized to
$1$, and move forward in time. We denote by $(p^{\alpha})$ the
momenta of the particles. The conservation of the quantity
$g_{\alpha\beta}p^{\alpha}p^{\beta}$ requires that the phase space
of the particle is the seven-dimensional submanifold
\begin{equation*}
     PM = \{g_{\alpha\beta}p^{\alpha}p^{\beta} = -1;\quad\
     p^{0}>0\}
\end{equation*}
 of $TM$ which is coordinatized by $(t,x^{i}, p^{i})$. If the coordinates
are such that the components $g_{0i}$ vanish then the component
$p^{0}$ is expressed by the other coordinates via
\begin{equation*}
 p^{0} = \sqrt{-g^{00}}\sqrt{1+g_{ij}p^{i}p^{j}}
 \end{equation*}
The distribution function of the particles is a non-negative
real-valued function denoted by $f$, that is defined on $PM$. In
addition we consider a scalar field $\phi$ which is a real-valued
function on $M$. The Einstein-Vlasov-scalar field system now
reads:
\begin{equation*}
\partial_{t}f + \frac{p^{i}}{p^{0}}\partial_{x^{i}}f -
\frac{1}{p^{0}}\Gamma_{\beta\gamma}^{i}p^{\beta}p^{\gamma}\partial_{p^{i}}f
 = 0
 \end{equation*}
\begin{equation*}
\nabla^\alpha\nabla_\alpha\phi=0
\end{equation*}
 \begin{equation*}
G_{\alpha\beta}  =  8 \pi T_{\alpha\beta}
\end{equation*}
\begin{equation*}
T_{\alpha\beta}  = -\int_{\mathbb{R}^{3}}fp_{\alpha}p_{\beta}\mid
g \mid^{\frac{1}{2}} \frac{dp^{1}dp^{2}dp^{3}}{p_{0}} +
(\nabla_{\alpha}\phi\nabla_{\beta}\phi -
\frac{1}{2}g_{\alpha\beta}\nabla_{\nu}\phi\nabla^{\nu}\phi)
\end{equation*}
where $p_{\alpha} = g_{\alpha\beta}p^{\beta}$,
 $|g|$ denotes the modulus of determinant of the metric $g_{\alpha\beta}$,
 $\Gamma_{\alpha\beta}^{\lambda}$
 the Christoffel symbols,
$G_{\alpha\beta}$ the Einstein tensor, and $T_{\alpha\beta}$ the
energy-momentum tensor.

Note that since the contribution of $f$ to the energy-momentum
tensor is divergence-free \cite {ehlers}, the form of the
contribution of the scalar field to the energy-momentum tensor
determines the field equation for $\phi$.

In \cite{rendall}, a definition of spacetimes with spherical,
plane and hyperbolic symmetry was given. The spacetime $(M,g)$ is
topologically of the form $]0, \infty[\times S^{1}\times S$, where
$S$ is a $2$-sphere, a $2$-torus, or a hyperbolic plane, in the
case of spherical, plane or hyperbolic symmetry respectively. We
now consider a solution of the Einstein-Vlasov-scalar field system
where all unknowns are invariant under one of these symmetries and
write the Einstein-Vlasov system in areal coordinates. The
 circumstances under which coordinates of this type exist are
 discussed in \cite{andreasson}. The metric $g$ takes the form
\begin{equation} \label{eq:1.1}
ds^{2} = - e^{2\mu(t,r)}dt^{2} + e^{2\lambda(t,r)}dr^{2} +
t^{2}(d\theta^{2} + \sin_{k}^{2}\theta d\varphi^{2})
\end{equation}
where
\begin{equation*}
 \sin_{k}\theta =
\begin{cases}
  \sin \theta  &\text{for $k = 1$  (spherical symmetry);}  \\
  1           &\text{for $k = 0$  (plane symmetry);}  \\
  \sinh\theta  &\text{for $k = -1$  (hyperbolic symmetry)}
\end{cases}
\end{equation*}
 $t > 0$ denotes a time-like coordinate, $r\in \mathbb{R}$ and
$(\theta, \varphi)$ range
 in the domains $[0,\pi]\times[0,2\pi]$, $[0,2\pi]\times[0,2\pi]$,
$[0,\infty[\times[0,2\pi]$
 respectively, and stand for angular coordinates. The functions $\lambda$
and $\mu$ are periodic in $r$ with period
 $1$. It has been shown in \cite{rein1} and \cite{rein} that due
 to the symmetry, $f$ can be written as a function of
 \begin{equation*}
 t, r, w := e^{\lambda}p^{1} \  {\rm and} \  F := t^{4}[(p^{2})^{2} +
\sin_k^2\theta (p^{3})^{2}],
 \end{equation*}
i.e. $f = f(t,r,w,F)$. In these variables, we have $p^{0} =
e^{-\mu} \sqrt{1+ w^{2}+ F/t^{2}}$. The scalar field is a function
of $t$ and $r$ which is periodic in $r$ with period 1.

We denote by a dot and by a prime the derivatives of the metric
components and of the scalar field with respect to $t$ and $r$
respectively. After calculation and using the results of
\cite{rein1}, we can write the complete Einstein-Vlasov-scalar
field system as follows:
\begin{equation} \label{eq:1.2}
\partial_{t}f + \frac{e^{\mu - \lambda}w}{\sqrt{1 + w^{2} +
F/t^{2}}} \partial_{r}f - (\dot{\lambda}w + e^{\mu - \lambda}
\mu'{\sqrt{1 + w^{2} + F/t^{2}}}) \partial _{w}f = 0
\end{equation}
\begin{equation} \label{eq:1.3}
e^{-2 \mu}(2t \dot{\lambda} + 1) + k = 8 \pi t^{2} \rho
\end{equation}
\begin{equation} \label{eq:1.4}
e^{-2\mu}(2t \dot{\mu} - 1) - k = 8 \pi t^{2}p
\end{equation}
\begin{equation} \label{eq:1.5}
\mu' = -4\pi t e^{\lambda + \mu}j
\end{equation}
\begin{equation} \label{eq:1.6}
e^{-2 \lambda}(\mu'' + \mu'(\mu' - \lambda')) - e^{-2
\mu}(\ddot{\lambda} +
(\dot{\lambda}+\frac{1}{t})(\dot{\lambda}-\dot{\mu})) = 4\pi q
\end{equation}
\begin{equation} \label{eq:1.7}
e^{-2\lambda} \phi'' - e^{-2\mu} \ddot{\phi} -
e^{-2\mu}(\dot{\lambda} - \dot{\mu} + \frac{2}{t})\dot{\phi} -
e^{-2\lambda}(\lambda' - \mu')
 \phi' = 0
\end{equation}
where (\ref{eq:1.7}) is the wave equation in $\phi$ and :
\begin{equation} \label{eq:1.8}
\begin{aligned}
 \rho(t,r)  = e^{-2\mu} T_{00}(t,r) &= \frac{\pi}{t^{2}}
\int_{-\infty}^{+\infty} \int_{0}^{+\infty} \sqrt{1 + w^{2} +
F/t^{2}} f(t,r,w,F)dFdw \\
 & \qquad + \frac{1}{2}(e^{-2\mu}
 \dot{\phi}^{2} +
e^{-2\lambda}{\phi'}^{2})
\end{aligned}
\end{equation}
\begin{equation} \label{eq:1.9}
\begin{aligned}
p(t,r)  = e^{-2\lambda} T_{11}(t,r) &= \frac{\pi}{t^{2}}
\int_{-\infty}^{+\infty}\int_{0}^{+\infty} \frac{w^2}{\sqrt{1
 + w^2 + F/t^{2}}} f(t,r,w,F)dFdw \\
 & \qquad + \frac{1}{2}(e^{-2\mu}
\dot{\phi}^{2} + e^{-2\lambda}{\phi'}^{2})
\end{aligned}
\end{equation}
\begin{equation} \label{eq:1.10}
j(t,r) = -e^{-(\lambda + \mu)} T_{01}(t,r) = \frac{\pi}{t^{2}}
\int_{-\infty}^{+\infty}\int_{0}^{+\infty}wf(t,r,w,F)dFdw
-e^{-(\lambda + \mu)} \dot{\phi} \phi'
\end{equation}
\begin{equation} \label{eq:1.11}
\begin{aligned}
q(t,r) &= \frac{2}{t^{2}} T_{22}(t,r) = \frac{2}{t^{2}\sin_k^2\theta} T_{33}(t,r,\theta)\\
  &= \frac{\pi}{t^{4}} \int
_{-\infty}^{\infty}\int_{0}^{\infty} \frac{F}{\sqrt{1 + w^{2}
 +F/{t^{2}}}} f(t,r,w,F)dFdw + e^{-2\mu} \dot{\phi}^{2} -
e^{-2\lambda}{\phi'}^{2}
\end{aligned}
\end{equation}
We are going to study the initial value problem corresponding to
this system and prescribe initial data at time $t=1$:
\begin{eqnarray}
&&f(1,r,w,F) = \overset{\circ}{f}(r,w,F),\ \  \lambda(1,r) =
\overset{\circ}{\lambda}(r),\ \  \mu(1,r) =
\overset{\circ}{\mu}(r),\nonumber      \\
&&\phi(1,r) = \overset{\circ}{\phi}(r),\ \ \dot{\phi}(1,r) =
\psi(r)\nonumber
\end{eqnarray}
The choice $t=1$ is made only for convenience. Analogous results hold
in the case that data are prescribed on any hypersurface $t=t_0>0$.

The paper is organized as follows. In section $2$, we split the
wave equation in $\phi$ into a system of two partial differential
equations of first order; we use it to bound the derivatives of
$\phi$ by the solutions of the field equations and we introduce an
auxiliary system. In section $3$, we solve each equation of the
auxiliary system when the other unknowns are fixed. Next, we
prove, using the results of \cite{rein1}, that under some
constraints on the initial data, the full system is equivalent to
the auxiliary system and reduces to a subsystem. Finally we solve
the constraint equation on the initial data. In section $4$, we
prove local in time existence and uniqueness of solutions in both
time directions by iterating the solutions of the auxiliary system
and we prove continuation criteria. In section 5 we show that
if $f=0$ global existence in the future holds when $k\le 0$.

\section{Auxiliary system and preliminary results}

 Using characteristic derivatives, we show in this section that
the first and second derivatives of $\phi$ can be bounded in terms
of $\lambda$ and $\mu$.
\begin{lemma} \label{L:2.1}
Let\ \ $D^{+} = e^{-\mu}\partial_{t} + e^{-\lambda}\partial_{r}$ \
; \quad
$D^{-} = e^{-\mu}\partial_{t} - e^{-\lambda}\partial_{r} \ $;\\
$ X =\dot{\phi}e^{-\mu} - \phi'e^{-\lambda}$  \ ; \quad $Y =
\dot{\phi}e^{-\mu} + \phi'e^{-\lambda}$ \ ;\\ $a
=(-\dot{\lambda}-\frac{1}{t})e^{-\mu} - \mu'e^{-\lambda}$ \ ;
\quad $b = -\frac{e^{-\mu}}{t}$ \ ; \quad $c =
(-\dot{\lambda}-\frac{1}{t})e^{-\mu} + \mu'e^{-\lambda}$ \\
 then $X$, $Y$ are solutions of the system
\begin{equation} \label{eq:2.1}
D^{+}X = aX + bY
\end{equation}
\begin{equation} \label{eq:2.2}
 D^{-}Y = bX + cY
\end{equation}
\end{lemma}
\textbf{Proof}: This results from a simple calculation, since
equation (\ref{eq:1.7}) is satisfied.$\square$

Note that the only derivatives of the metric coefficients which
appear in these equations are $\dot\lambda$ and $\mu'$ and that
these are the same ones which appear in the Vlasov equation. This
fact plays an important role in the proof of the local existence
theorem.

The full system above is overdetermined and we will show that a
solution $(f,\lambda,\mu,\phi)$ of the subsystem consisting of
equations (\ref{eq:1.2}), (\ref{eq:1.3}), (\ref{eq:1.4}) and
(\ref{eq:1.7}) solves the remaining equations (\ref{eq:1.5}) and
(\ref{eq:1.6}). Notice that such a solution determines the right
hand side of (\ref{eq:1.5}) which is then a given function say
$\tilde{\mu}$. Then, since (\ref{eq:1.4}) already provides $\mu$,
an idea introduced in \cite{rein1}, and that we will follow here,
is to replace $\mu'$ in (\ref{eq:1.2}) and (\ref{eq:1.7})  by an
auxiliary function $\tilde{\mu}$, which is not assumed a priori to
be a derivative, and prove later that, under certain conditions,
$\tilde{\mu}$ is nothing else than $\mu'$. We then introduce the
following auxiliary system obtained by coupling
(\ref{eq:1.3})-(\ref{eq:1.4}) to the equations obtained by
replacing $\mu'$ by $\tilde{\mu}$ in (\ref{eq:1.2}),(\ref{eq:1.5})
and (\ref{eq:2.1})-(\ref{eq:2.2}), i.e :
\begin{equation} \label{eq:2.3}
\partial_{t}f + \frac{e^{\mu - \lambda}w}{\sqrt{1 + w^{2} +
F/t^{2}}} \partial_{r}f - (\dot{\lambda}w + e^{\mu - \lambda}
\widetilde{\mu}{\sqrt{1 + w^{2} + F/t^{2}}}) \partial _{w}f = 0
\end{equation}
\begin{equation} \label{eq:2.4}
\tilde{\mu} = -4\pi t e^{\lambda + \mu}j
\end{equation}
and
\begin{equation} \label{eq:2.5}
D^{+}X = \tilde{a}X + bY
\end{equation}
\begin{equation} \label{eq:2.6}
 D^{-}Y = bX + \tilde{c}Y
\end{equation}
 Where $\mu'$ is substituted by $\widetilde{\mu}$ in $a$ and $c$ to
obtain $\widetilde{a}$ and $\widetilde{c}$ respectively. We first
specify the regularity properties which we require.
\begin{definition}\label{D:2.2}
Let $I \subseteqq ]0, \infty[$ be an interval and $(t,r) \in
I\times\mathbb{R}$.\\
 a)\ \   $f\in C^{1}(I \times \mathbb{R}^{2} \times [0, \infty[)$ is
 regular if $f(t,r+1,w,F)$ = $f(t,r,w,F)$ for $(t,r,w,F) \in I \times
\mathbb{R}^{2}
 \times [0,
 \infty[$, $f \geq 0$ and $\supp f(t,r,.,.)$ is compact uniformly in
 $r$ and locally uniformly in $t$.\\
 b)\ \ $\mu\in C^{1}(I \times \mathbb{R})$ is regular, if
$\mu' \in C^{1}(I \times
 \mathbb{R})$ and $\mu(t,r+1)$ = $\mu(t,r)$.\\
 c) \  \ $\lambda \in C^{1}(I \times \mathbb{R})$ is regular, if $\dot{\lambda}
  \in C^{1}(I \times
 \mathbb{R})$ and $\lambda(t,r+1)$ = $\lambda(t, r)$.\\
 d)\ \ $\widetilde{\mu}$  (or $\phi_{1}$, $\phi_{2}$) $\in C^{1}
(I \times \mathbb{R})$ is regular,
 if $\widetilde{\mu}(t,r+1)$ = $\widetilde{\mu}(t,r)$.\\
 e)\ \ $\rho $ (or $p$, $j$, $q$) $\in C^{1}(I \times \mathbb{R})$ is regular,
 if $\rho(t,r+1)$ = $\rho(t,r)$.\\
 f)\ \  $\phi \in C^{2}(I \times \mathbb{R})$ is regular, if $\phi(t,r+1)$ =
 $\phi(t,r)$.
\end{definition}
    Let us estimate the first and second order
 derivatives of the scalar field $\phi$, using the characteristic
 curves of (\ref{eq:1.7}) and equations (\ref{eq:2.5}) and (\ref{eq:2.6}).
\begin{proposition} \label{P:2.3}
Let
\begin{equation*}
K_{0} = 2 \sup \{\mid \psi(r) \mid e^{-\overset{\circ} {\mu}(r)} +
{\mid \overset{\circ}{\phi}'(r) \mid} e^{-\overset{\circ}
{\lambda}(r)} \ ; \ r \in \mathbb{R}\}
\end{equation*}
\begin{equation*}
m(t) =  \sup \{\mid \dot{\lambda}(t,r) \mid + \frac{2}{t} + {\mid
\tilde{\mu}(t,r) \mid} e^{(\mu - \lambda)(t,r)} \ ; \ r \in
\mathbb{R}\}
\end{equation*}
\begin{equation*}
K(t) =  \sup \{(|X|^2+|Y|^2)^{\frac{1}{2}}(t,r) \ ; \ r \in
\mathbb{R}\}
\end{equation*}
then if $(X,Y)$ is a solution of (\ref{eq:2.5}) and (\ref{eq:2.6})
with
\begin{equation}
X(1)=e^{-\overset{\circ} {\mu}(r)}\psi(r)-e^{-\overset{\circ}
{\lambda}(r)}\overset{\circ}{\phi}'(r)
\end{equation}
and
\begin{equation}
Y(1)=e^{-\overset{\circ} {\mu}(r)}\psi(r)+e^{-\overset{\circ}
{\lambda}(r)}\overset{\circ}{\phi}'(r)
\end{equation}
then:

1) If $t \in ]T,1]$, $T \geq 0$, we have
\begin{equation}\label{eq:2.7}
K(t) \leq K_{0} + 3\int_{t}^{1} m(s) K(s)ds
\end{equation}
2) If $t \geq 1$ the analogous estimate holds with the limits $t$
and $1$ exchanged in the integral in (\ref{eq:2.7}).
\end{proposition}
\textbf{Proof}: The characteristic curves $(t,\gamma_{i})$, $i =
1, 2 $ of the second order partial differential equation
(\ref{eq:1.7}) satisfy the differential equation $\dot{\gamma_{i}}
= \pm e^{\mu-\lambda}$. On these characteristic curves, we have
from Lemma $2.1$ , $D^{+} = D^{-} = e^{-\mu}\frac{d}{dt}$. Then
(\ref{eq:2.5})-(\ref{eq:2.6})  become
\begin{equation*}
\begin{cases}
\frac{d}{dt}X(t,\gamma_{1}(t)) =  e^{\mu}(\tilde{a}X +
b Y)(t,\gamma_{1}(t))\\
\frac{d}{dt}Y(t,\gamma_{2}(t)) =  e^{\mu}(bX +
\tilde{c}Y)(t,\gamma_{2}(t))
\end{cases}
\end{equation*}
Integrate this system on $[t,1]$ (or on $[1,t]$), take the
absolute value in each equation, add the two inequalities and take
the supremum of each term to obtain (\ref{eq:2.7}). $\square$

    Next, we obtain by direct calculation, using
(\ref{eq:2.5})-(\ref{eq:2.6}), the following result:
\begin{lemma} \label{L:2.3}
Let $D^{+}$ and $D^{-}$ be defined as in Lemma $2.1$ and
define \\ $X_{1} = e^{-\lambda} \partial_{r}X$,\ \ \  $Y_{1} = e^{-\lambda} \partial_{r}Y$ ;\\
$ b_{1} = (-2\dot{\lambda} - \frac{1}{t})e^{-\mu} - (\tilde{\mu} +
\mu')e^{-\lambda}$ \ ; \quad $ b_{2} = \frac{-e^{-\mu}}{t}$ \ ;\\
 $b_{3} = -\dot{\lambda}'e^{-\mu-\lambda} + (\lambda'\tilde{\mu} - \tilde{\mu}\mu' -
 \tilde{\mu}')e^{-2\lambda}$ \ ;\quad
$b_{4} = (-2\dot{\lambda} - \frac{1}{t})e^{-\mu} + (\tilde{\mu} +
\mu')e^{-\lambda}$ \ ; \\
 $b_{5} = -\dot{\lambda}'e^{-\mu-\lambda} - (\lambda'\tilde{\mu} - \tilde{\mu}\mu' -
 \tilde{\mu}')e^{-2\lambda}$ \ ;\

\noindent If $X$ and $Y$ satisfy (\ref{eq:2.5}) and (\ref{eq:2.6})
then $X_{1}$ and $Y_{1}$ satisfy
\begin{equation} \label{eq:2.8}
D^{+}X_{1} = b_{1}X_{1} + b_{2}Y_{1} + b_{3}X
\end{equation}
\begin{equation} \label{eq:2.9}
 D^{-}Y_{1}= b_{2}X_{1} + b_{4}Y_{1} + b_{5}Y
\end{equation}
\end{lemma}
\begin{proposition} \label{P:2.4}
Let $K(t)$ be defined as in Proposition $2.3$ and set:
\begin{align*}
A_{0} &= 2 \sup \{[(\mid \psi' \mid + \mid \overset{\circ}{\mu}'
\mid \mid \psi \mid)e^{-\overset{\circ}{\mu}-\overset{\circ}{\lambda}} + (\mid
\overset{\circ}{\phi}'' \mid + \mid \overset{\circ}{\lambda}' \mid
\mid\overset{\circ}{\phi}'\mid)e^{-2\overset{\circ}{\lambda}}](r) \
; \
r \in \mathbb{R}\} \\
A(t) &=  \sup \{[X_1^2+Y_1^2](t,r) \ ; \ r \in
\mathbb{R}\}\\
 v(t) &= \sup\{\frac{2}{t} + 2\mid \dot{\lambda}\mid
+ (\mid \tilde{\mu}\mid + \mid \mu' \mid)e^{\mu - \lambda}(t,r) \ ; \ r\in \mathbb{R}\}\\
h(t) &= \sup \{[\mid \dot{\lambda}'\mid e^{-\lambda} + (\mid \mu'
\mid \mid \tilde{\mu} \mid + \mid \lambda'\mid \mid
\tilde{\mu}\mid + \mid
\tilde{\mu'} \mid) e^{\mu-2\lambda}](t,r) \ ; \  r \in \mathbb{R} \}\\
\end{align*}
If in addition to the assumptions of Proposition 2.3 the
quantities $X_1$ and $Y_1$ satisfy (\ref{eq:2.8}) and
(\ref{eq:2.9}) and agree with $e^{-\lambda}\partial_{r}X$
and $e^{-\lambda}\partial_{r}Y$ respectively for $t=1$ then \\
\textbf{1)} If $t \in ]T,1]$, $T \geq 0$, we have the estimate
\begin{equation} \label{eq:2.10}
A(t) \leq A_{0} + 3\int_{t}^{1}(v(s)A(s) + h(s)K(s))ds
\end{equation}
\textbf{2)} If $t \geq 1$ the analogous estimate holds with the
limits $t$ and $1$ exchanged in the integral in (\ref{eq:2.10}).
\end{proposition}
\textbf{Proof}: Analogous to the proof of Proposition $2.3$, using
this time
Lemma $2.4$.$\square$\\
Note that the factor $e^{-\lambda}$ in the definition of $X_1$ and
$Y_1$ is very important. Without it the above derivation would not
work since the derivative $\lambda'$ would occur in $b_1$ and $b_4$.

\section{The reduced system}
We first solve each equation of the auxiliary system introduced in
section $2$, when the other unknowns are fixed (in the form to be
used later for the iteration). In order to clarify our statements,
we introduce the notations $\phi_{1}$, $\phi_{2}$ in place of
$\dot{\phi}$, $\phi'$.

\begin{proposition} \label{P:3.1}
    \textbf{1)} Let $\bar{f}$, $\bar{\lambda}$, $\bar{\mu}$, $\tilde{\mu}$,
$\bar{\phi}_{1}$, $\bar{\phi}_{2}$ be regular for $(t,r) \in
I\times \mathbb{R}$ , $I \subset ]0, \infty[$. Substitute $f$,
$\lambda$, $\mu$,  $\dot{\phi}$, $\phi'$ respectively by
$\bar{f}$, $\bar{\lambda}$, $\bar{\mu}$, $\bar{\phi}_{1}$,
$\bar{\phi}_{2}$  in $\rho$ and $p$ to define $\bar{\rho}$ and
$\bar{p}$. Suppose that $1 \in I$, $\overset{\circ}{f}$ $\in
C^{1}(\mathbb{R}^{2} \times [0, \infty))$,
$\overset{\circ}{\lambda}$, $\overset{\circ}{\mu}$ $\in
C^{1}(\mathbb{R})$ and are periodic of period $1$ in $r$. Assume
that:
\begin{equation} \label{eq:3.1}
\frac{e^{-2\overset{\circ}{\mu}(r)}+k}{t} - k + \frac{8\pi}{t}
\int_{t}^{1}s^{2}\bar{p}(s,r)ds > 0 , \ \ (t,r) \in I \times
\mathbb{R}
\end{equation}
then the system
\begin{equation} \label{eq:3.2}
\partial_{t}f + \frac{e^{\bar{\mu} - \bar{\lambda}}w}{\sqrt{1 + w^{2} +
F/t^{2}}} \partial_{r}f - (\dot{\bar{\lambda}}w + e^{\bar{\mu} -
\bar{\lambda}} \tilde{\mu}{\sqrt{1 + w^{2} + F/t^{2}}})
\partial _{w}f = 0
\end{equation}
\begin{equation} \label{eq:3.3}
e^{-2 \mu}(2t \dot{\lambda} + 1) + k = 8 \pi t^{2}\bar{\rho}
\end{equation}
\begin{equation} \label{eq:3.4}
e^{-2 \mu}(2t \dot{\mu} - 1) - k = 8 \pi t^{2}\bar{p}
\end{equation}
has a unique regular solution $(f,\lambda,\mu)$ on $ I \times
\mathbb{R}$ with $f(1) = \overset{\circ}{f}$, $\lambda(1) =
\overset{\circ}{\lambda}$, $\mu(1) = \overset{\circ}{\mu}$. This
solution is given by
\begin{equation} \label{eq:3.5}
 f(t,r,w,F) = \overset{\circ}{f}((R,W)(1,t,r,w,F),F)
\end{equation}
where $(R,W)$ is the solution of the characteristic system:
\begin{equation} \label{eq:3.6}
\frac{d}{ds}(r,w) =  (\frac{e^{\bar{\mu} - \bar{\lambda}}w}{\sqrt{1
+ w^{2} + F/t^{2}}}, -\dot{\bar{\lambda}}w - e^{\bar{\mu} -
\bar{\lambda}} \tilde{\mu}{\sqrt{1 + w^{2} + F/t^{2}}})
\end{equation}
satisfying $(R,W)(t,t,r,w,F) = (r,w)$;
\begin{equation} \label{eq:3.7}
e^{-2\mu(t,r)} =  \frac{e^{-2\overset{\circ}{\mu}(r)} + k}{t} - k
+ \frac{8\pi}{t} \int_{t}^{1}s^{2}\bar{p}(s,r)ds
\end{equation}
\begin{equation} \label{eq:3.8}
\dot{\lambda}(t,r) = 4\pi te^{2\mu}\bar{\rho}(t,r) - \frac{1 +
ke^{2\mu}(t,r)}{2t}
\end{equation}
\begin{equation} \label{eq:3.9}
\lambda(t,r) = \overset{\circ}{\lambda}(r) -
\int_{t}^{1}\dot{\lambda}(s,r)ds
\end{equation}
If $I = ]T, 1]$ (respectively $I = [1,T[$) with $T \in [0,1[$
(respectively $T \in ]1, \infty[$), then there exists some $T^{*}
\in [T, 1]$ (respectively $T^{*} \in ]1, T]$) such that condition
(\ref{eq:3.1}) holds on $]T^{*},1] \times \mathbb{R}$
(respectively $[1,T^{*}[ \times \mathbb{R}$). $T^{*}$ is
independent of $\bar{p}$ if $I = ]T,1]$, whereas it depends on
$\bar{p}$ if $I = [1,T[$.

    \textbf{2)} Let $\lambda$, $\mu$, $\bar{\lambda}$, $\bar{\mu}$ be regular;
let $C$, $D$ be regular as $\tilde{\mu}$ (see definition $2.2$).
Set
\begin{equation} \label{eq:3.10}
X = \phi_{1}e^{-\mu} - \phi_{2}e^{-\lambda} \ \ ,\ \ Y =
\phi_{1}e^{-\mu} + \phi_{2}e^{-\lambda}
\end{equation}
and define the operators $\bar{D^{+}}$, $\bar{D^{-}}$ as $D^{+}$,
$D^{-}$ in Lemma $2.1$, with $\lambda$, $\mu$ substituted
respectively by $\bar{\lambda}$, $\bar{\mu}$. Assume that $\psi$
$\in C^{1}(\mathbb{R})$, $\overset{\circ}{\phi}$ $\in
C^{2}(\mathbb{R})$ are periodic of period $1$. Then the system
\begin{equation} \label{eq:3.11}
\bar{D}^{+}X = C
\end{equation}
\begin{equation} \label{eq:3.12}
\bar{D}^{-}Y = D
\end{equation}
has a unique regular solution ($\phi_{1}, \phi_{2}$) such that
$(\phi_{1}, \phi_{2})(1) = (\psi,\overset{\circ}{\phi}')$.
\end{proposition}
\textbf{Proof}:
    1) the proof of this point is the same as that of Proposition $2.4$ in
\cite{rein1}.
The only thing added is the existence of $T^{*}$ we now prove and
that will replace in the case of local existence, the hypothesis
$\overset{\circ}{\mu} \leq 0$ if $k = -1$ in \cite{rein1}.

If $I = ]T,1]$; since $\bar{p} \geq 0$, the left hand side of
(\ref{eq:3.1}) is bounded from below by $h(t,r) =
\frac{e^{-2\overset{\circ}{\mu}(r)}+k}{t} - k$. If $k \in\{0,1\}$,
we have, since $\frac{1}{t} \geq 1$, $h(t,r)\geq
e^{-2\overset{\circ}{\mu}(r)}> 0$ and we can take $T^{*} = T$. If
$k = -1$, since $\overset{\circ}{\mu}$ is bounded, there exists
$\beta > 0$, such that $h(1,r) = e^{-2\overset{\circ}{\mu}(r)} >
\beta$. By the continuity of $t\mapsto h(t,r)$ at $t=1$, we
conclude that:
\begin{equation}\label{eq:3.13}
\exists \ T^{*}\ \in \ ]T,1] \ \ {\rm such\ that} \ \
e^{-2\mu(t,r)} \geq \ \ h(t,r)> \frac{\beta}{2},\ \ t\in ]T^{*},1]
\ \ {\rm and} \ \ \bar{p} \geq 0.
\end{equation}
If $I = [1,T[$, define $h(t,r)$ to be all the left hand side of
(\ref{eq:3.1}) and proceed as above in the case $k = -1$, to
obtain $T^{*}$ that depends this time on $\bar{p}$.

2) The system (\ref{eq:3.11})-(\ref{eq:3.12}) in $(X,Y)$ is a
first order linear hyperbolic system, and the existence of a
unique solution with the prescribed data $(X,Y)(1) =
(e^{-\overset{\circ}{\mu}}\psi -
e^{-\overset{\circ}{\lambda}}\overset{\circ}{\phi}',
e^{-\overset{\circ}{\mu}}\psi +
e^{-\overset{\circ}{\lambda}}\overset{\circ}{\phi}')$ is given by
a theorem of Friedrichs \cite{friedrichs}. (See also the paper of
Douglis \cite{douglis} where existence of $C^1$ solutions of
hyperbolic systems in one space dimension was proved for $C^1$
initial data in the more general quasilinear case.) We then deduce
from the relations (\ref{eq:3.10}) that define a bijection $(X,Y)
\mapsto (\phi_{1},\phi_{2})$, the existence of a unique regular
solution $(\phi_{1},\phi_{2})$ of (\ref{eq:3.11})-(\ref{eq:3.12})
such that $(\phi_{1},\phi_{2})(1) =
(\psi,\overset{\circ}{\phi}')$. This completes the proof of Proposition
$3.1$.$\square$ \\

In fact, since the equations for $X$ and $Y$ are decoupled, it is
not necessary to use existence results for hyperbolic systems in
the above proof; it suffices to solve a parameter-dependent ordinary
differential equation. The above procedure has the advantage that it
can easily be generalized to problems where the corresponding equations
are coupled as they are, for instance, in the case of a nonlinear
scalar field.

Now we show that the solution of the subsystem consisting of
equations
 (\ref{eq:1.2}), (\ref{eq:1.3}), (\ref{eq:1.4}) and (\ref{eq:1.7}) also
 satisfies equations (\ref{eq:1.5}) and (\ref{eq:1.6}), and that the auxiliary system is equivalent to the full system.
 \begin{proposition} \label{P:3.2}

 \textbf{1)} Let $(f,\lambda,\mu,\phi)$ be a regular solution of
 (\ref{eq:1.2}), (\ref{eq:1.3}), (\ref{eq:1.4}) and (\ref{eq:1.7}) on some
 time interval $I \in ]0, \infty[$ with $1 \in I$, and let the
 initial data satisfy (\ref{eq:1.5}) for $t = 1$ with
$\overset{\circ}{\phi}, \psi \in
 C^{1}(\mathbb{R})$, in particular $\overset{\circ}{\mu} \in
 C^{2}(\mathbb{R})$. Then (\ref{eq:1.5}) and (\ref{eq:1.6}) hold for
 all $t \in I$, in particular $\mu \in C^{2}(I \times
 \mathbb{R})$.

\textbf{2)} Let $(f,\lambda,\mu,\tilde{\mu},\phi_{1},\phi_{2})$ be
a regular solution of equations (\ref{eq:2.3}),(\ref{eq:1.3}),
(\ref{eq:1.4}), (\ref{eq:2.4}), (\ref{eq:2.5}) and (\ref{eq:2.6})
with the initial data $(f,\lambda,\mu,\phi_{1},\phi_{2})(1) =
(\overset{\circ}{f},\overset{\circ}{\lambda},
\overset{\circ}{\mu},\psi,\overset{\circ}{\phi}')$ that are as in
proposition \ref{P:3.1} and satisfy (\ref{eq:1.5}) for $t=1$. Then
there exists a unique regular function $\phi$ satisfying $\phi(1)
= \overset{\circ}{\phi}$, $\dot{\phi}(1) = \psi$ such that
$(f,\lambda,\mu,\phi)$ solves the full system
(\ref{eq:1.2})-(\ref{eq:1.11}). The function $\phi$ is given by:
$\phi(t,r)= \overset{\circ}{\phi}(r) + \int_{1}^{t}
\phi_{1}(s,r)ds$.
\end{proposition}

\textbf{Proof}: 1) Firstly, we prove that (\ref{eq:1.5}) holds.
From equations (\ref{eq:1.9}), (\ref{eq:1.2}), (\ref{eq:1.7}) and
integration by parts, it follows that
\begin{equation}\label{eq:3.14}
\begin{aligned}
\int_{1}^{t} p'(s,r)s^{2}ds = \pi \int_{1}^{t}
\int_{-\infty}^{\infty} \int_{0}^{\infty} \frac{w^{2}}{\sqrt{1 +
w^{2}+F/{s^{2}}}}
\partial_{r} f(s,r,w,F) dFdwds \\ \qquad
+ \int_{1}^{t}(-\mu'\dot{\phi}^{2}e^{-2\mu} -
\lambda'{\phi'}^{2}e^{-2\lambda} + \dot{\phi}'\dot{\phi}e^{-2\mu}
+ \phi'\phi''e^{-2\lambda})s^{2}ds \\ \qquad =
[-e^{\lambda-\mu}j(s,r)s^{2}]_{s = 1}^{s = t} - \int
_{1}^{t}(\dot{\lambda}+ \dot{\mu})e^{\lambda - \mu}j(s,r)s^{2}ds -
\int_{1}^{t}\mu'(\rho + p)(s,r)s^{2} ds
\end{aligned}
\end{equation}
From equation (\ref{eq:1.4}), we obtain
\begin{equation*}
e^{-2\mu(t,r)} =  \frac{e^{-2\overset{\circ}{\mu}(r)} + k}{t} - k
+ \frac{8\pi}{t} \int_{t}^{1}s^{2}p(s,r)ds
\end{equation*}
 and differentiating this with respect to $r$, yields:
\begin{equation}\label{eq:3.15}
t\mu'e^{-2\mu} = \overset{\circ}{\mu}'e^{-2\overset{\circ}{\mu}} +
4\pi \int_{1}^{t}s^{2}p'(s,r) ds
\end{equation}
Substituting for the last integral by (\ref{eq:3.14}), using (\ref{eq:1.3})
and (\ref{eq:1.4})
to replace $\dot\lambda+\dot\mu$ and assuming the fact that the
constraint equation (\ref{eq:1.5}) holds for $t =1$, relation (\ref{eq:3.15})
implies
\begin{equation*}
-te^{2\mu}(\mu' + 4 \pi e^{\lambda + \mu}jt) = 4\pi
\int_{1}^{t}(\rho + p)s^{2}(\mu' + 4 \pi e^{\lambda + \mu}js) ds
\end{equation*}
and since the left hand side is zero at $t=1$, we obtain
\begin{equation*}
\mu' + 4 \pi te^{\lambda + \mu}j = 0
\end{equation*}
on $I$, i.e (\ref{eq:1.5}) holds for all $t \in I$. In particular
this relation shows that $\mu$ is $C^{2}$ with respect to $r$ with
\begin{equation*}
\mu'' = (\lambda' + \mu')\mu' - 4 \pi te^{\lambda + \mu}j'.
\end{equation*}
Now we prove that equation (\ref{eq:1.6}) holds. From
(\ref{eq:1.10}), (\ref{eq:1.2}), (\ref{eq:1.7}) and integration by
parts we obtain the identity
\begin{align*}
j'(t,r) & = \frac{\pi}{t^{2}} e^{\lambda - \mu}
\int_{-\infty}^{\infty} \int_{0}^{\infty}[-\sqrt{1+w^{2} +
F/{t^{2}}}\partial_{t}f \\
& + (\dot{\lambda}w \sqrt{1+w^{2}+F/{t^{2}}} + e^{\mu -
\lambda}\mu'(1+w^{2} +
F/{t^{2}}))\partial_{w}f]dFdw\\
& + 2\mu' e^{-(\lambda+\mu)}\dot{\phi}\phi' -
e^{-(\lambda+\mu)}\dot{\phi}'\phi' -
e^{\lambda-3\mu}[\dot{\phi}\ddot{\phi}+(\dot{\lambda}-\dot{\mu}+\frac{2}{t})
\dot{\phi}^{2}]\\
& = -\frac{\pi}{t^{2}} e^{\lambda - \mu} \int_{-\infty}^{\infty}
\int_{0}^{\infty} \sqrt{1+w^{2} + F/{t^{2}}}\partial_{t}f dFdw -
\dot{\lambda}e^{\lambda - \mu}(\rho + p -
\dot{\phi}^{2}e^{-2\mu}\\
& - {\phi'}^{2}e^{-2\lambda}) - 2\mu'j - e^{-(\lambda +
\mu)}\dot{\phi}'\phi' - e^{-3\mu + \lambda}(\dot{\phi}\ddot{\phi}
- (\dot{\lambda} - \dot{\mu} + \frac{2}{t})\dot{\phi}^{2}) \ .
\end{align*}
Since by (\ref{eq:1.8})
\begin{align*}
\dot{\rho}(t,r) & =
-2\frac{\pi}{t^{3}}\int_{-\infty}^{\infty}\int_{0}^{\infty}\sqrt{1+w^{2}+
F/{t^{2}}}f dFdw \\
& +
\frac{\pi}{t^{2}}\int_{-\infty}^{\infty}\int_{0}^{\infty}\sqrt{1+w^{2}
+ F/{t^{2}}}\partial_{t}f
dFdw\\
&+ \frac{\pi}{t^{2}}\int_{-\infty}^{\infty}\int_{0}^{\infty}(-\frac{F}{t^{3}})
  \frac{f}{\sqrt{1+w^{2}+ F/{t^{2}}}}dFdw\\
&+ e^{-2\mu}(-\dot{\mu}\dot{\phi}^{2} + \dot{\phi}\ddot{\phi}) +
e^{-2\lambda}(-\dot{\lambda}{\phi'}^{2} + \dot{\phi}'\phi')\\
& = \frac{-2\rho}{t} - \frac{q}{t} +
\frac{2e^{-2\mu}}{t}\dot{\phi}^{2} +
e^{-2\mu}(-\dot{\mu}\dot{\phi}^{2} + \dot{\phi}\ddot{\phi}) +
e^{-2\lambda}(-\dot{\lambda}{\phi'}^{2} + \dot{\phi}'\phi')\\
& +
\frac{\pi}{t^{2}}\int_{-\infty}^{\infty}\int_{0}^{\infty}\sqrt{1+w^{2}
+ F/{t^{2}}}\partial_{t}fdFdw \ ;
\end{align*}
From (\ref{eq:1.3}) we obtain
\begin{equation*}
\dot{\lambda}(t,r) = 4\pi te^{2\mu}\rho(t,r) - \frac{1 +
ke^{2\mu}(t,r)}{2t}
\end{equation*}
and differentiating with respect to $t$ yields:
\begin{align*}
\ddot{\lambda} &= 4\pi e^{2\mu}\rho + 8\pi t\dot{\mu} e^{2\mu}\rho
+ \frac{1+ke^{2\mu}}{2t^{2}} + 4\pi te^{2\mu}\dot{\rho}-\frac{k\dot{\mu}e^{2\mu}}{t}\\
& = \frac{\dot{\mu}-\dot{\lambda}}{t} - 4\pi qe^{2\mu} +
2\dot{\lambda}\dot{\mu} + 8\pi \dot{\phi}^{2} + 4\pi
t(-\dot{\mu}\dot{\phi}^{2} + \dot{\phi}\ddot{\phi}) + 4\pi
te^{2\mu - 2\lambda}(-\dot{\lambda}{\phi'}^{2} + \dot{\phi}'\phi')\\
& +
4\frac{\pi^{2}}{t}e^{2\mu}\int_{-\infty}^{\infty}\int_{0}^{\infty}\sqrt{1+w^{2}
+ F/{t^{2}}}\partial_{t}f dFdw
\end{align*}
Combining all these relations gives the remaining field equation
(\ref{eq:1.6}).

2) subtract the two equations
(\ref{eq:2.5})-(\ref{eq:2.6}) to obtain
\begin{equation} \label{eq:3.16}
\dot{\phi}_{2} - \phi_{1}' = (\tilde{\mu} - \mu')\phi_{1}
\end{equation}
Let us first prove that $\mu' = \tilde{\mu}$. Consider equation
(\ref{eq:3.15}) and write $p=p_1+p_2$ where $p_1$ and $p_2$ are
the contributions to $p$ made by $f$ and $\phi$ respectively. We
obtain from \cite{rein1} that studies the case $\phi = 0$, and in
which $p$ corresponds to $p_{1}$ here, that we have, using
definitions (\ref{eq:1.8}), (\ref{eq:1.9}) and (\ref{eq:1.10}) of
$\rho$, $p$, $j$ :
\begin{align*}
\int_{1}^{t}s^{2}p_{1}'(s,r) ds &= - \int_{1}^{t}(\dot{\lambda} +
\dot{\mu})e^{\lambda - \mu}s^{2}(j + e^{-\lambda -
\mu}\phi_{1}\phi_{2}) ds - [e^{\lambda - \mu}s^{2}(j + e^{-\lambda
- \mu}\phi_{1}\phi_{2})]_{1}^{t} \\
& -  \int_{1}^{t} \tilde{\mu}[(\rho+p) -
(e^{-2\mu}\phi_{1}^{2}+e^{-2\lambda}\phi_{2}^{2})]s^{2} ds
\end{align*}
Now we obtain, using integration by parts and (\ref{eq:3.16})
that:
\begin{align*}
\int_{1}^{t}s^{2}p_{2}'(s,r)ds =
[s^{2}e^{-2\mu}\phi_{1}\phi_{2}]_{1}^{t} +
\int_{1}^{t}[(\dot{\lambda} + \dot{\mu})\phi_{1}\phi_{2}e^{-2\mu}
- \tilde{\mu}(\phi_{1}^{2}e^{-2\mu} + \phi_{2}^{2}e^{-2\lambda})]
s^{2}ds
\end{align*}
Adding (\ref{eq:1.3}) and (\ref{eq:1.4}) yields
\begin{equation} \label{eq:3.17}
\dot{\lambda} + \dot{\mu} = 4 \pi te^{2\mu}(\rho + p)
\end{equation}
 Then, if we use (\ref{eq:3.17}) and the definition (\ref{eq:2.4}) of
 $\tilde{\mu}$, we obtain from (\ref{eq:3.15}) :
 \begin{equation*}
t \mu'e^{-2\mu} = e^{-2\overset{\circ}{\mu}}(\overset{\circ}{\mu}'
+ 4 \pi
e^{\overset{\circ}{\lambda}+\overset{\circ}{\mu}}\overset{\circ}{j})+
t \tilde{\mu}e^{-2\mu} \ \ {\rm for \ all} \ \ t \in I \subset
]0,\infty[,
\end{equation*}
Hence, if (\ref{eq:1.5}) holds for $t = 1$, then, $t \mu'e^{-2\mu}
= t \tilde{\mu}e^{-2\mu}$ and $\mu' = \tilde{\mu}$.\\

    Now we prove the existence of $\phi$. Define $\phi$ by :
$\phi(t,r) = \overset{\circ}{\phi}(r) + \int_{1}^{t} \phi_{1}(s,r)
ds$. Then $\phi(1) = \overset{\circ}{\phi}$ , $\dot{\phi}(1) =
\phi_{1}(1) = \psi$ and $\dot{\phi} = \phi_{1}$. Now
(\ref{eq:3.16}) implies, since $\mu' = \tilde{\mu}$, that
$\dot{\phi_{2}} = \phi_{1}'$, hence the relation $\phi_{2}(1) =
\overset{\circ}{\phi}'$ implies $\phi' = \phi_{2}$.\\ The relation
$\mu' = \tilde{\mu}$ also implies that the systems
(\ref{eq:2.1})-(\ref{eq:2.2}) and (\ref{eq:2.5})-(\ref{eq:2.6})
are identical. Then a direct calculation, using the fact that
$(\phi_{1},\phi_{2})$ satisfies the system
(\ref{eq:2.1})-(\ref{eq:2.2}) shows that $\phi$ satisfies
(\ref{eq:1.7}).
$\square$ \\

We conclude this section with a proposition dealing with the
solvability of the constraint equation (\ref{eq:1.5}) for $t = 1$.
Let $\tilde\psi=e^{-\mu}\psi$.
\begin{proposition} \label{P:3.4}
Given a function $\overset{\circ}{\lambda}(r)$, a non-negative
function $\bar f(r,w,F)$ and functions $\overset{\circ}{\phi}(r)$
and $\tilde\psi(r)$, all periodic in $r$ and regular, there exists a
function $\overset{\circ}{\mu}(r)$, periodic in $r$ and regular,
such that the constraint equation
\begin{equation*}
\overset{\circ}{\mu}' = -4\pi
 e^{\overset{\circ}{\lambda} +
 \overset{\circ}{\mu}}\overset{\circ}{j}
\end{equation*}
holds for a non-negative function $\overset{\circ}{f}$. It can be assumed
that $\overset{\circ}{f}=\bar f+a\Phi$, where $\Phi(r,w,F)$ is a fixed
function, independent of the particular choice of input data, and $a$ is a
suitable constant.
\end{proposition}
\textbf{Proof}: This can be proved just as in \cite{tchapnda1},
with $\Phi$ chosen as in that reference.\ $\square$

This result shows that it is possible to produce a plentiful supply
of initial data. It cannot be applied to produce data with $f=0$.
A way of doing that is to adjust $\tilde\psi$ instead of adjusting
$f$.

\section{Local existence and continuation of solutions}
In this section we prove using an iteration the local existence
and uniqueness of solutions of the Einstein-Vlasov-scalar field
system together with continuation criteria. Let us first use the
solution $(f,\lambda,\mu,\tilde{\mu},\phi_{1},\phi_{2})$ of the
auxiliary system consisting of the equations (\ref{eq:2.3}),
(\ref{eq:1.3}), (\ref{eq:1.4}), (\ref{eq:2.4}), (\ref{eq:2.5}) and
(\ref{eq:2.6}), to construct a sequence of iterative solutions as
follows.
 Define $\overset{\circ}{\tilde{\mu}} :=
\overset{\circ}{\mu}'$, $\lambda_{0}(t,r) :=
\overset{\circ}{\lambda}(r)$, $\mu_{0}(t,r) :=
\overset{\circ}{\mu}(r)$, $\tilde{\mu}_{0}(t,r) :=
\overset{\circ}{\tilde{\mu}}(r)$,
$g_0(t,r)=\psi(r)$, $h_0(t,r)=\overset{\circ}\phi'$
for $t \in ]0, 1]$, $r \in
\mathbb{R}$. If $\lambda_{n-1}$, $\mu_{n-1}$, $\tilde{\mu}_{n-1}$
are already defined and regular on $]T^{*}, 1] \times \mathbb{R}$
then let
\begin{equation} \label{eq:4.1}
  G_{n-1}(t, r, w, F) :=
\left(\frac{w
e^{\mu_{n-1}-\lambda_{n-1}}}{\sqrt{1+w^{2}+F/t^{2}}},
-\dot{\lambda}_{n-1}w -
e^{\mu_{n-1}-\lambda_{n-1}}\tilde{\mu}_{n-1}
\sqrt{1+w^{2}+F/t^{2}}\right)
\end{equation}
 and denote by $(R_{n}, W_{n})(s, t, r, w, F)$ the solution of the
characteristic system
\begin{eqnarray*}
\frac{d}{ds}(R, W) = G_{n-1}(s, R, W, F)
\end{eqnarray*}
with initial data
\begin{eqnarray*}
(R_{n}, W_{n})(t,t,r,w,F) = (r, w); \ \ (t, r, w, F)\in ]0,1]
\times \mathbb{R}^{2} \times [0, \infty[ \ ;
\end{eqnarray*}
note that $F$ is constant along characteristics. Define
\begin{equation}\label{eq:4.2}
f_{n}(t, r, w, F) := \overset{\circ}{f}\left((R_{n}, W_{n})(1, t,
r, w, F), F\right),
\end{equation}
that is, $f_{n}$ is the solution of
\begin{equation} \label{eq:4.3}
\partial_{t}f_{n} + \frac{w
e^{\mu_{n-1}-\lambda_{n-1}}}{\sqrt{1+w^{2}+F/t^{2}}}\partial_{r}f_{n}
- (\dot{\lambda}_{n-1}w +
e^{\mu_{n-1}-\lambda_{n-1}}\tilde{\mu}_{n-1}\sqrt{1+w^{2}+F/t^{2}})
\partial_{w}f_{n}
= 0
\end{equation}
with $f_{n}(1) = \overset{\circ}{f}$. Define $\rho_{n}$, $p_{n}$,
$j_{n}$, $q_{n}$ by the formulas
(\ref{eq:1.8}), (\ref{eq:1.9}), (\ref{eq:1.10}) and (\ref{eq:1.11}) with
$f$, $\lambda$, $\mu$, $\dot{\phi}$, $\phi'$ respectively replaced
by $f_{n}$, $\lambda_{n-1}$, $\mu_{n-1}$,  $g_{n-1}$, $h_{n-1}$,
$(n \geq 1)$. Using Proposition $3.1$, $1)$, define $\mu_{n}$ and
$\lambda_{n}$ to be the solutions of
\begin{equation}\label{eq:4.4}
e^{-2\mu_{n}(t,r)} = \frac{e^{-2\overset{\circ}{\mu}(r)}+k}{t} - k
+ \frac{8\pi}{t}\int_{t}^{1}s^{2}p_{n}(s,r)ds
\end{equation}
\begin{equation}\label{eq:4.5}
\dot{\lambda}_{n}(t,r) = 4\pi te^{2\mu_{n}(t,r)}\rho_{n}(t,r) -
\frac{1+ke^{2\mu_{n}}}{2t}
\end{equation}
\begin{equation}\label{eq:4.6}
\lambda_{n}(t,r) = \overset{\circ}{\lambda}(r) -
\int_{t}^{1}\dot{\lambda}_{n}(s,r)ds
\end{equation}
and set
\begin{equation}\label{eq:4.7}
\tilde{\mu}_{n}(t,r) = -4\pi te^{(\mu_{n} +
\lambda_{n})(t,r)}j_{n}(t,r)
\end{equation}
Notice that, by Proposition $3.1$, the right hand side of
(\ref{eq:4.4}) is positive on $]T^{*}, 1]$, $\forall n$.
Now define $g_{n}$ and $h_{n}$ using Proposition $3.1$, $2)$
to satisfy the conditions that the quantities
\begin{equation*}
X_{n} = e^{-\mu_{n}} g_{n} - e^{-\lambda_{n}} h_{n} , \ \ Y_{n} =
e^{-\mu_{n}} g_{n} + e^{-\lambda_{n}} h_{n}
\end{equation*}
are solutions of the system
\begin{equation} \label{eq:4.8}
D^{+}_{n-1}X_{n} = a_{n-1}X_{n-1} + b_{n-1}Y_{n-1}
\end{equation}
\begin{equation} \label{eq:4.9}
D^{-}_{n-1}Y_{n} = b_{n-1}X_{n-1} + c_{n-1}Y_{n-1}
\end{equation}
where $D^{+}_{n-1}$, $D^{-}_{n-1}$, $a_{n-1}$, $b_{n-1}$ and
$c_{n-1}$ are defined in the same way as $D^{+}$, $D^{-}$,
$\tilde{a}$, $b$, $\tilde{c}$ (see Lemma $2.1$), with $\mu$,
$\lambda$, $\dot{\phi}$, $\phi'$, $\tilde{a}$, $b$, $\tilde{c}$
substituted respectively by $\mu_{n-1}$, $\lambda_{n-1}$,
$g_{n-1}$, $h_{n-1}$, $a_{n-1}$, $b_{n-1}$, $c_{n-1}$. Now $K_{0}$
and $A_{0}$ being defined in Propositions $2.3$ and $2.5$, we
introduce the following quantities that are similar to those
defined in those propositions:
\begin{equation*}
K_{n}(t) =  \sup \{(g_{n}^{2} e^{-2\mu_{n}} +
h_{n}^{2}e^{-2\lambda_{n}})^{\frac{1}{2}}(t,r) \ ; \ r \in
\mathbb{R}\}
\end{equation*}
\begin{equation}\label{eq:4.10}
\begin{cases}
\begin{aligned}
A_{n}(t) &=  \sup \{e^{-2\lambda_{n}}[(g_{n}' - \mu_{n}'
g_{n})^{2}e^{-2\mu_{n}} + (h_{n}' -
\lambda_{n}'h_{n})^{2}e^{-2\lambda_{n}}](t,r) \ ; \ r \in
\mathbb{R}\} \\
m_{n-1}(t) &= \sup\{\frac{2}{t} + (\mid \dot{\lambda}_{n-1}\mid +
\mid \tilde{\mu}_{n-1}\mid e^{\mu_{n-1} - \lambda_{n-1}})(t,r) \ ;
\ r\in \mathbb{R}\} \\
 v_{n-1}(t) &= \sup\{\frac{2}{t} + 2\mid \dot{\lambda}_{n-1}\mid
+ (\mid \tilde{\mu}_{n-1}\mid + \mid \mu'_{n-1} \mid )e^{\mu_{n-1} - \lambda_{n-1}}(t,r) \ ; \ r\in \mathbb{R}\} \\
\beta_{n-1}(t) &= \sup \{[\mid \dot{\lambda}'_{n-1}\mid
e^{-\lambda_{n-1}} + (\mid \mu'_{n-1} \mid \mid \tilde{\mu}_{n-1}
\mid + \mid \lambda'_{n-1}\mid \mid \tilde{\mu}_{n-1}\mid \\
& \ \qquad \ \ \qquad \ \ \qquad \ \qquad \ \qquad \ + \mid
\tilde{\mu}'_{n-1} \mid)e^{\mu_{n-1} - 2\lambda_{n-1}}](t,r) \ ; \
r \in \mathbb{R} \}
\end{aligned}
\end{cases}
\end{equation}
Now we proceed for (\ref{eq:4.8})-(\ref{eq:4.9}) the same way as
we did for (\ref{eq:2.5})-(\ref{eq:2.6}) to establish the
inequality (\ref{eq:2.7}) and we obtain the following analogous
inequality:
\begin{equation} \label{eq:4.11}
K_{n}(t) \leq K_{0} + 3\int_{t}^{1} m_{n-1}(s) K_{n-1}(s)ds
\end{equation}
We can use (\ref{eq:4.8})-(\ref{eq:4.9}) to establish a system for
$(e^{-\lambda_{n-1}}\partial_{r}X_{n}, e^{-\lambda_{n-1}}\partial_{r}Y_{n})$
analogous to (\ref{eq:2.8})-(\ref{eq:2.9}) from which we deduce the
following inequality which is analogous to (\ref{eq:2.10})
\begin{equation} \label{eq:4.12}
A_{n}(t) \leq A_{0} + 3\int_{t}^{1}(v_{n-1}(s)A_{n-1}(s) +
\beta_{n-1}(s)K_{n-1}(s))ds
\end{equation}
Throughout the paper, $  \parallel . \parallel$ denotes the
$L^{\infty}$-norm on the function space in question; we use the fact
that by (\ref{eq:4.2}), $\parallel f_{n}(t) \parallel$ = $
\parallel \overset{\circ}{f} \parallel$ for $n \in \mathbb{N}$ and
$t \in ]T^{*}, 1]$. The numerical constant $C$ may change from
line to line and does not depend on $n$ or $t$ or the initial data.
In order to prove the local existence theorem, we prove
respectively in the next two propositions: \\
 - a uniform bound on the momenta in the support of distribution functions
$f_{n}$, and a uniform bound of the first derivatives with respect to $r$ of
the functions $f_{n}$,
 $\lambda_{n}$, $\mu_{n}$, $g_{n}$, $h_{n}$;\\
 - the convergence of the iterates.
\begin{proposition} \label{P:4.1}
We take $\overset{\circ}{f}$ as in proposition $3.1$ and such that
\begin{equation}\label{eq:4.13}
\supp \overset{\circ}{f} \subset [0, W_{0}] \times [0,F_{0}], \
W_{0}>0, \ F_{0}>0.
\end{equation}
 then there exist nonnegative constants $T_{1}, T_{2}$ such that
 the
quantities
\begin{equation*}
Q_{n}(t) = \sup\{\mid w \mid; \ (r,w,F)\in \supp f_{n}(t)\} \ for
\ all \ t \in [T_{1}, 1].
\end{equation*}
\begin{equation*}
B_{n}(t) = \sup\{\parallel \partial_{r}f_{n}(s)\parallel +
A_{n-1}(s); \ t\leq s \leq 1 \} \ for \ all \ t \in [T_{2}, 1].
\end{equation*}
and $K_{n}(t)$ are uniformly bounded in n.
\end{proposition}
\textbf{Proof}:  Firstly we bound $Q_{n}(t)$ and $K_{n}(t)$.
On $\supp f_{n}(t)$, we have
\begin{equation}\label{eq:4.14}
\sqrt{1 + w^{2}+ F/{t^{2}}} \leq \sqrt{1 + Q_{n}^{2}+
F_{0}/{t^{2}}} \leq \frac{1}{t}(1 + F_{0})(1 + Q_{n}(t))
\end{equation}
and thus
\begin{equation*}
\parallel \rho_{n}(t) \parallel, \parallel p_{n}(t) \parallel, \parallel j_{n}(t)\parallel \leq
\frac{C}{t^{3}}(1 + F_{0})^{2}(1 + Q_{n}(t))^{2}
\parallel \overset{\circ}{f}\parallel
 + (K_{n-1}(t))^{2}
\end{equation*}
{}From (\ref{eq:3.13}), we have, setting $C_{0} =
\frac{\beta}{2}$,
\begin{equation} \label{eq:4.15}
e^{-2\mu_{n}}(t,r) \geq \frac{C_{0}}{t}
\end{equation}
Hence, proceeding as in step $1$ of the proof of theorem $3.1$ in
\cite{rein1} and using the expression for $j_{n}$, we have :
\begin{equation} \label{eq:4.16}
Q_{n+1}(t) \leq W_{0} + C_{1}\int_{t}^{1}\frac{1}{s}(1 +
Q_{n}(s))^{2}(1 + K_{n-1}(s))^{2}(1 +  Q_{n+1}(s))ds
\end{equation}
with $C_{1} = \frac{C}{C_{0}}[(1+F_{0})^{2} ( 1+
\parallel\overset{\circ}{f} \parallel)$.\
Next we have, using (\ref{eq:4.5})-(\ref{eq:4.7}) and
(\ref{eq:4.15})
\begin{equation}\label{eq:4.17}
\mid \dot{\lambda_{n}}(s,r)\mid + \mid
(\tilde{\mu}_{n}e^{\mu_{n}-\lambda_{n}})(s,r)\mid \leq
C_{1}\frac{(1 + Q_{n}(s))^{2}}{s}(1 + K_{n-1}(s))^{2}
\end{equation}
we then deduce from (\ref{eq:4.11}) that
\begin{equation}\label{eq:4.18}
K_{n+1}(t) \leq K_{0} + C_{1}\int_{t}^{1}\frac{(1 +
Q_{n}(s))^{2}}{s}(1 + K_{n-1}(s))^{2}K_{n}(s)ds
\end{equation}
Add (\ref{eq:4.16}) and (\ref{eq:4.18}) to obtain
\begin{equation} \label{eq:4.19}
Q_{n+1}(t) + K_{n+1}(t) \leq W_{0} + K_{0} +
C_{1}\int_{t}^{1}\frac{1}{s}(1 + Q_{n}(s))^{2}(1 +
K_{n-1}(s))^{2}(1 +  Q_{n+1}(s) + K_{n}(s))ds
\end{equation}
Now define $H_{n}(t):= \sup \{Q_{m}(t) + K_{m}(t) \ ; \ m \leq n
\}$. $(H_{n})_{n \in \mathbb{N}}$ is an increasing sequence. Then,
use (\ref{eq:4.19}) and the inequalities obtained by replacing in
(\ref{eq:4.19}), $n$ by any $m \leq n$, to obtain :
\begin{equation*}
 H_{n+1}(t) \leq W_{0} + K_{0} +
C_{1}\int_{t}^{1}\frac{1}{s}(1 + H_{n+1}(s))^{5}ds
\end{equation*}
Let $z_{1}$ be the left maximal solution of the equation
\begin{equation*}
 z_{1}(t) = W_{0} + K_{0} +
C_{1}\int_{t}^{1}\frac{1}{s}(1 + z_{1}(s))^{5}ds
\end{equation*}
which exists on some interval $]T_{1},1]$ with $T_{1} \in
[T^{*},1[$. By comparing the solution of the integral inequality with that
of the corresponding integral equation it follows that
\begin{equation*}
 H_{n+1}(t) \leq z_{1}(t), \ t \in ]T_{1},1[, \ n \in \mathbb{N}.
\end{equation*}
Since $Q_{n}(t) + K_{n}(t) \leq H_{n+1}(t)$, we obtain
\begin{equation*}
K_{n}(t) , \  Q_{n}(t) \leq z_{1}(t), \ t \in ]T_{1},1[, \ n \in
\mathbb{N}.
\end{equation*}
 And all the quantities which were estimated against $Q_{n}$ and
 $K_{n}$ in the above argument are bounded by certain powers of
 $z_{1}$ on $]T_{1},1]$. Namely there exists a continuous function $C_{2}(t)$
which depends only on $z_{1}$ as an increasing function, such that
\begin{equation}\label{eq:4.20}
\begin{cases}
\begin{aligned}
\parallel \mu_{n}(t)\parallel,  \  \parallel \lambda_{n}(t)\parallel, \
\parallel \dot{\lambda}_{n}(t)\parallel, \  \parallel \rho_{n}(t)\parallel, \
 \parallel p_{n}(t)\parallel, \\ \parallel j_{n}(t)\parallel,
 \parallel \tilde{\mu}_{n}e^{\mu_{n}-\lambda_{n}}
\parallel, \ \parallel g_{n}(t)\parallel, \ \parallel h_{n}(t)\parallel
  \ \leq C_{2}(t)
\end{aligned}
\end{cases}
\end{equation}
    Now we bound $B_{n}(t)$. Following step $2$ of the proof of
theorem $3.1$ in \cite{rein1}, we have, using (\ref{eq:4.20}), the
estimates
\begin{equation} \label{eq:4.21}
\parallel \rho_{n}'(t)\parallel,  \  \parallel p_{n}'(t)\parallel, \
\parallel j_{n}'(t)\parallel,\parallel \mu_{n}'(t)\parallel, \  \parallel
\dot{\lambda}_{n}'(t)\parallel, \
\parallel \lambda_{n}'(t)\parallel  \ \leq C_{2}(t)(C_{3} +
B_{n}(t))
\end{equation}
\begin{equation} \label{eq:4.22}
\parallel \tilde{\mu}_{n}'e^{\mu_{n}-\lambda_{n}}
\parallel \leq C_{2}(t)(C_{3} +
B_{n}(t))
\end{equation}
\begin{equation} \label{eq:4.23}
\parallel \partial_{r}f_{n+1}(t) \parallel \leq
\parallel \partial_{(r,w)}\overset{\circ}{f} \parallel
 \exp \int_{t}^{1} C_{2}(s)(C_{3} +
B_{n}(s))ds
\end{equation}
with $C_{3} = \parallel \overset{\circ}\lambda' \parallel +
\parallel
\overset{\circ}\mu'e^{-2\overset{\circ}\mu} \parallel + 1$. We use
(\ref{eq:4.20}), (\ref{eq:4.21}), (\ref{eq:4.22}) and relation
(\ref{eq:4.12}) to obtain
\begin{equation*}
A_{n+1}(t) \leq A_{0} + \int_{t}^{1}C_{2}(s)(C_{3} + B_{n}(s))
A_{n}(s)ds
\end{equation*}
Let $D_{n}(t):= \sup\{A_{m}(t)| m \leq n \}$ and $E_{n}(t):=
\sup\{B_{m}(t)| m \leq n \}$. $\{D_{n}\}$ and $\{E_{n}\}$ are
increasing sequences. Therefore
\begin{equation}\label{eq:4.24}
A_{n+1}(t) \leq A_{0} + \int_{t}^{1}C_{2}(s)(C_{3} + E_{n}(s))
D_{n+1}(s)ds
\end{equation}
then we deduce by replacing $n$ by any $m \leq n$ in
(\ref{eq:4.24})
\begin{equation*}
D_{n+1}(t) \leq A_{0} + \int_{1}^{t}C_{2}(s)(C_{3} + E_{n}(s))
D_{n+1}(s)ds
\end{equation*}
Which gives:
\begin{equation}\label{eq:4.25}
D_{n+1}(t) \leq  2A_{0}\exp \int_{t}^{1}C_{2}(s)(C_{3} +
E_{n}(s))ds
\end{equation}
Now add (\ref{eq:4.23})-(\ref{eq:4.25}) to obtain
\begin{equation*}
B_{n+1}(t)  \leq (2A_{0}+\parallel
\partial_{r,w}\overset{\circ}{f} \parallel) \exp \int_{t}^{1}C_{2}(s)(C_{3} + E_{n}(s))ds
\end{equation*}
and deduce by replacing $n$ by every $m \leq n$ that:
\begin{equation*}
E_{n+1}(t) \leq C_{4}\exp \int_{t}^{1}C_{2}(s)(C_{3} +
E_{n+1}(s))ds
\end{equation*}
where $C_{4} = 2A_{0} + \parallel
\partial_{(r,w)}\overset{\circ}{f} \parallel$.
Let $z_{2}$ be the left maximal solution of
\begin{equation*}
z_{2}(t) = C_{4}\exp \int_{t}^{1}C_{2}(s)(C_{3} + z_{2}(s))ds
\end{equation*}
i.e
\begin{equation*}
\dot{z_{2}}(t) = - C_{2}(t)(C_{3} + z_{2}(t))z_{2}(t), \ \ \
z_{2}(1) = C_{4};
\end{equation*}
which exists on an interval $]T_{2}, 1] \subset ]T_{1}, 1]$. Then
we have
\begin{equation*}
E_{n+1}(t) \leq  z_{2}(t), \ t \in ]T_{2}, 1], \ n \in \mathbb{N}
\end{equation*}
and so
\begin{equation*}
A_{n}(t) \ ,  B_{n}(t) \leq  z_{2}(t), \ t \in ]T_{2}, 1], \ n \in
\mathbb{N}
\end{equation*}
and all the quantities estimated against $B_{n}$ above can be
bounded in terms of $z_{2}$ on $]T_{2}, 1]$, uniformly in
$n$.$\square$
\begin{remark} \label{r:4.2}
The sequences $\lambda_{n}$, $\mu_{n}$, $f_{n}$,
$\tilde{\mu}_{n}e^{\mu_{n}- \lambda_{n}}$, $\rho_{n}$, $p_{n}$,
$j_{n}$, $g_{n}$, $h_{n}$, $\lambda_{n}'$, $\mu_{n}'$, $f_{n}'$,
$g_{n}', h_{n}'$, $\dot{\lambda}_{n}$, $\dot{\mu}_{n}$,
$\dot{g_{n}}$, $\dot{h_{n}}$, $\dot{\lambda_{n}}'$, $\rho_{n}'$,
$p_{n}'$, $j_{n}'$, $\tilde{\mu_{n}}'$, are uniformly bounded in
the $L^{\infty}-$norm by a function of $t$ on $[T_{1}^{*}, 1]$
with $T_{1}^{*} = \max (T_{1}, T_{2})$.
\end{remark}

     In order to prove the convergence of the
iterates in the following proposition, we introduce auxiliary
variables $\tilde{g}_{n}$ and $\tilde{h}_{n}$ defined by
$\tilde{g}_{n} = g_{n}e^{-\mu_{n}}$,\\ $\tilde{h}_{n} =
h_{n}e^{-\lambda_{n}}$, \ for $n \in \mathbb{N}$.
\begin{proposition} \label{P:4.3}
Let $[T_{3}, 1] \subset [T_{2}, 1]$, be an arbitrary compact
subset on which the previous estimates hold. Then on such an
interval, the iterates converge uniformly.
\end{proposition}
\textbf{Proof}: Define for $t \in [T_{3}, 1]$:
\begin{align*}
\alpha_{n}(t) &:= \sup\{\parallel (f_{n+1} - f_{n})(s)\parallel +
\parallel (\tilde{g}_{n+1} - \tilde{g}_{n})(s)\parallel  +
\parallel (\tilde{h}_{n+1} - \tilde{h}_{n})(s)\parallel \\
& +
 \parallel (\lambda_{n+1} - \lambda_{n})(s)\parallel + \parallel (\mu_{n+1} - \mu_{n})(s)\parallel
; \ \ t \leq s \leq 1 \}
\end{align*}
and let $C$ denote a constant which may depend on the functions
$z_{1}$ and $z_{2}$ introduced previously. If we consider the new
quantities
\begin{equation*}
\tilde{X}_{n} = (\tilde{g}_{n+1} - \tilde{g}_{n}) -
(\tilde{h}_{n+1} - \tilde{h}_{n}) ;\ \ \ \tilde{Y}_{n} =
(\tilde{g}_{n+1} - \tilde{g}_{n}) + (\tilde{h}_{n+1} -
\tilde{h}_{n}),
\end{equation*}
then we obtain by subtracting the system
(\ref{eq:4.8})-(\ref{eq:4.9}) written for $n+1$ and $n$, the new
system
\begin{equation}\label{eq:4.26}
D_{n}^{+}\tilde{X}_{n} = a_{n}\tilde{X}_{n-1} +
b_{n}\tilde{Y}_{n-1} + F_{n}
\end{equation}
\begin{equation}\label{eq:4.27}
D_{n}^{-}\tilde{Y}_{n} = b_{n}\tilde{X}_{n-1} +
c_{n}\tilde{Y}_{n-1} + G_{n}
\end{equation}
where
\begin{align*}
F_{n} &= (a_{n} - a_{n-1} + b_{n} - b_{n-1})\tilde{g}_{n-1} +
(a_{n-1} - a_{n} + b_{n} - b_{n-1})\tilde{h}_{n-1}\\
& + (e^{-\mu_{n-1}} - e^{-\mu_{n}})(\dot{\tilde{g}}_{n} -
\dot{\tilde{h}}_{n}) + (e^{-\lambda_{n-1}} -
e^{-\lambda_{n}})(\tilde{g}'_{n} - \tilde{h}'_{n})
\end{align*}
and substitute in $F_{n}$, $\tilde{h}_{n}'$ and $\tilde{g}_{n}'$
respectively by $-\tilde{h}_{n}'$ and $-\tilde{g}_{n}'$ to obtain
$G_{n}$.\\
Now let
\begin{equation*}
\theta_{n}(t) = \sup\{\mid \tilde{g}_{n+1} -\tilde{g}_{n}\mid +
\mid \tilde{h}_{n+1} -\tilde{h}_{n}\mid ; \ r \in \mathbb{R} \}
\end{equation*}
Thus similarly to (\ref{eq:4.11}), we have :
\begin{equation} \label{eq:4.28}
\theta_{n}(t) \leq 3\int_{t}^{1}(m_{n-1}(s)\theta_{n-1}(s) +
\sup\{\mid F_{n}(s,r)\mid + \mid G_{n}(s,r)\mid ; \ r\in
\mathbb{R}\})ds
\end{equation}
Using the mean value theorem to express the differences
$e^{-\mu_{n}} - e^{-\mu_{n-1}}$, \\ $e^{-\lambda_{n}} -
e^{-\lambda_{n-1}}$ and remark $4.2$, then (\ref{eq:4.28}) gives
\begin{equation} \label{eq:4.29}
\mid \tilde{g}_{n+1} - \tilde{g}_{n}\mid
 + \mid \tilde{h}_{n+1} - \tilde{h}_{n}\mid \leq C\int_{t}^{1}(\alpha_{n-1} +
\mid \tilde{\mu}_{n} -\tilde{\mu}_{n-1}\mid
 + \mid \dot{\lambda}_{n}
-\dot{\lambda}_{n-1}\mid)(s)ds
\end{equation}
 The expressions of $\rho_{n}$, $p_{n}$, $j_{n}$ yield, using
 Proposition $4.1$, that
\begin{align*}
\mid\rho_{n+1} - \rho_{n}\mid(t), \mid p_{n+1} - p_{n}\mid(t),
\mid j_{n+1} - j_{n}\mid(t) \leq C \alpha_{n}(t)
\end{align*}
{}From (\ref{eq:4.5}) and (\ref{eq:4.7}),  we have respectively
\begin{equation}\label{eq:4.30}
\mid\dot{\lambda}_{n} - \dot{\lambda}_{n-1}\mid(t)  \leq C
\alpha_{n-1}(t)
\end{equation}
\begin{equation}\label{eq:4.31}
 \mid \tilde{\mu}_{n} - \tilde{\mu}_{n-1}\mid(t) \leq
C \alpha_{n-1}(t)
\end{equation}
Using the two previous inequalities, (\ref{eq:4.29}) gives
\begin{equation} \label{eq:4.32}
(\mid\tilde{g}_{n+1} - \tilde{g}_{n}\mid
 + \mid \tilde{h}_{n+1} - \tilde{h}_{n}\mid)(t) \leq
 C\int_{t}^{1}\alpha_{n-1}(s)ds
\end{equation}
By the mean value theorem, (\ref{eq:4.4}) gives :
\begin{equation} \label{eq:4.33}
\mid \mu_{n+1} - \mu_{n}\mid(t) \leq C\int_{t}^{1} \alpha_{n}(s)
ds
\end{equation}
(\ref{eq:4.6}) gives :
\begin{equation} \label{eq:4.34}
\mid \lambda_{n+1} - \lambda_{n}\mid(t) \leq
C\int_{t}^{1}\alpha_{n}(s)ds
\end{equation}
Following step $3$ in the proof of theorem $3.1$ in \cite{rein1},
we have
\begin{equation*}
\mid (R,W)_{n+1} - (R,W)_{n} \mid(1,t,r,w,F) \leq
C\int_{t}^{1}\alpha_{n-1}(s)ds
\end{equation*}
 This implies using (\ref{eq:4.2}) and the mean value theorem
\begin{equation} \label{eq:4.35}
\mid (f_{n+1} - f_{n})(t) \mid  \leq  C\mid
\partial_{r,w}\overset{\circ}{f} \mid
\int_{t}^{1}\alpha_{n-1}(s)ds
\end{equation}
Adding (\ref{eq:4.32}), (\ref{eq:4.33}), (\ref{eq:4.34}),
(\ref{eq:4.35}), then
\begin{equation*}
\alpha_{n}(t) \leq C\int_{t}^{1}(\alpha_{n}(s) +
\alpha_{n-1}(s))ds \ ; \ \ n \geq 1.
\end{equation*}
By Gronwall's inequality  \ \  $\alpha_{n}(t) \ \leq
C\int_{t}^{1}\alpha_{n-1}(s)ds$ ;\\ and by induction
\begin{equation*}
\alpha_{n}(t) \leq C^{n+1}\frac{(1-t)^{n}}{n!} \leq
\frac{C^{n+1}}{n!} \ \ for \ \ n \in \mathbb{N}, \ t \in [T_{3},
1]
\end{equation*}
Since the series $\sum \frac{C^{n+1}}{n!}$ converges, we deduce
the convergence of $\sum \alpha_{n}$ which implies that
$\alpha_{n}\rightarrow 0$ for $n \rightarrow \infty$. Every
difference term which appears in $\alpha_{n}$, converges to zero.
We deduce the uniform convergence of
\begin{equation}\label{eq:4.36}
 f_{n}, \lambda_{n}, \mu_{n},
\tilde{g}_{n}, \tilde{h}_{n}, \dot{\lambda}_{n}, \dot{\mu}_{n},
\tilde{\mu}_{n}, \rho_{n}, p_{n}, j_{n}.
\end{equation}
And in $L^{\infty}-$norm, $\lambda_{n}\rightarrow \lambda$; \
$\mu_{n}\rightarrow \mu$; \ $\tilde{\mu}_{n}\rightarrow
\tilde{\mu}$; \ $f_{n}\rightarrow f$; \ $\tilde{g}_{n}\rightarrow
\tilde{g}$; \ $\tilde{h}_{n}\rightarrow \tilde{h}$.$\square$

     It remains to show that the
limits $\tilde{g}, \tilde{h}, f, \lambda, \mu $ are $C^{1}$, $f$
solves the Vlasov equation (\ref{eq:1.2}), $ \lambda, \mu $ solve
the field equations (\ref{eq:1.3})-(\ref{eq:1.4}), and to show the
existence of function $\phi$ that solves the wave equation
(\ref{eq:1.7}). This is the subject of the next theorem.
\begin{theorem} \label{T:4.4}{\bf(local existence)}
Let
 $\overset{\circ}{f} \in C^{1}(\mathbb{R}^{2} \times [0, \infty[)$
with \\ $\overset{\circ}{f}(r+1,w,F) = \overset{\circ}{f}(r,w,F)$
for $(r,w,F) \in \mathbb{R}^{2} \times [0, \infty[$,
 $\overset{\circ}{f}\geq 0$, and
\begin{eqnarray*}
W_{0} := \sup \{ |w| | (r,w,F) \in {\rm supp} \overset{\circ}{f}
\} <
 \infty
\end{eqnarray*}
\begin{eqnarray*}
F_{0} := \sup \{ F | (r,w,F) \in {\rm supp} \overset{\circ}{f} \}
< \infty
\end{eqnarray*}
Let $\overset{\circ}{\lambda}, \psi \in C^{1}(\mathbb{R})$,
$\overset{\circ}{\mu}, \overset{\circ}{\phi} \in
C^{2}(\mathbb{R})$ with $\overset{\circ}{\lambda}(r) =
\overset{\circ}{\lambda}(r+1)$, $\overset{\circ}{\mu}(r) =
\overset{\circ}{\mu}(r+1)$,\\ $\overset{\circ}{\phi}(r) =
\overset{\circ}{\phi}(r+1)$  and
\begin{eqnarray*}
\overset{\circ}{\mu}'(r) =
 -4 \pi e^{\overset{\circ}{\lambda} +
\overset{\circ}{\mu}}\overset{\circ}{j}(r) , \ \  r \in \mathbb{R}
\end{eqnarray*}
Then there exists a unique, left maximal, regular solution $(f,
\lambda, \mu, \phi)$ of system (\ref{eq:1.2})-(\ref{eq:1.11}) with
$(f, \lambda, \mu, \phi)(1) = (\overset{\circ}{f},
\overset{\circ}{\lambda}, \overset{\circ}{\mu},
\overset{\circ}{\phi})$ and $\dot{\phi}(1) = \psi$ on a time
interval $]T, 1]$ with $T \in [0, 1[$.
\end{theorem}
{\bf Proof} : Consider the sequences of iterates constructed at
the begining of this section and the limit obtained in the above
proposition. We need the uniform convergence of the
derivatives of these iterates.\\
In what follows, we fix $T_{4} \in [T_{2},1]$, $t \in [T_{4},1]$,
$|w| < U$, $F < F_{0}$, $t \leq s \leq 1$.\\
\underline{Step $1$}: Convergence of $(\partial_{r}
f_{n})$ and $(\partial_{w} f_{n})$.\\
Following step $4$ in the proof of theorem $3.1$ in \cite{rein1},
and using (\ref{eq:4.36}), we can establish with minor changes,
using Proposition $4.1$, that if we set:
\begin{equation}\label{eq:4.37}
\xi_{n}(s) = e^{(\lambda_{n} - \mu_{n})(s,r)}\partial R_{n}(s, t,
r, w,F)
\end{equation}
\begin{equation}\label{eq:4.38}
\eta_{n}(s) = \partial W_{n}(s)
  + (\sqrt{1+w^{2}+F/s^{2}} e^{\lambda_{n} - \mu_{n}}
\dot{\lambda}_{n})
\partial R_{n}(s)
\end{equation}
in which $\partial$ stands for $\partial_{r}$ or $\partial_{w}$
and $s \mapsto (R_{n}(s),W_{n}(s))$ the indicated solution of the
characteristic system associated to equation (\ref{eq:4.3}) in
$f_{n}$, and then: $\forall \epsilon > 0$, $\exists N \in
\mathbb{N}$ such that we have, for $n > N$:
\begin{equation}\label{eq:4.39}
(\mid \xi_{n+1} - \xi_{n}\mid + \mid \eta_{n+1} -
 \eta_{n}\mid)(s) \leq C \epsilon + C \int_{s}^{1}(\mid \xi_{n+1} -
\xi_{n}\mid +  \mid \eta_{n+1} - \eta_{n}\mid)(\tau) d\tau
\end{equation}
in which $C>0$ stands, as in what follows, for a constant that may
change from line to line. (\ref{eq:4.39}) implies by Gronwall's
lemma, that $(\xi_{n})$ and $(\eta_{n})$ converge uniformly. Now,
 since the transformation $(\partial R_{n}, \partial W_{n}) \mapsto (\xi_{n},
 \eta_{n})$ defined by (\ref{eq:4.37})-(\ref{eq:4.38}) is invertible
 with convergent coefficients, this implies the convergence of
 $\partial_{r,w}(R_{n},W_{n})$ and, given (\ref{eq:4.2}), the
 convergence of $(\partial_{r}
f_{n})$ and $(\partial_{w} f_{n})$.\\
\underline{Step $2$}: convergence of $(\lambda_{n}'), (\mu_{n}'),
(\tilde{\mu}_{n}'), (\tilde{g}_{n}'), (\tilde{h}_{n}')$.
  We set
\begin{equation}\label{eq:4.40}
\begin{aligned}
\gamma_{n}(t) = \sup \{ |\xi_{n+1}- \xi_{n}|(s) + |\eta_{n+1}-
\eta_{n}|(s) + \parallel (\mu_{n+1}' - \mu_{n}')(s)
\parallel \\ + \parallel (\lambda_{n+1}' - \lambda_{n}')(s)
\parallel\ + \parallel (\tilde{g}_{n+1}' - \tilde{g}_{n}')(s)
\parallel\ + \parallel (\tilde{h}_{n+1}' - \tilde{h}_{n}')(s)
\parallel\ ; t \leq s \leq 1 \}
\end{aligned}
\end{equation}
Now since $(\mu_{n})$, $(\tilde{\mu}_{n})$, $(\dot{\lambda}_{n})$,
$(\tilde{g}_{n})$, $(\tilde{h}_{n})$, $(\rho_{n})$, $(j_{n})$
converge uniformly, we take the above integer $N$ sufficiently
large so that we have for $n > N$:
\begin{equation}\label{eq:4.41}
\begin{aligned}
\parallel (\mu_{n+1}-\mu_{n})(s)\parallel ,\parallel
(j_{n+1}-j_{n})(s)
\parallel,
\parallel (\tilde{\mu}_{n}-\tilde{\mu}_{n-1})(s)\parallel,
\parallel( \dot{\lambda}_{n+1}-\dot{\lambda}_{n})(s) \parallel,\\
\parallel (\tilde{g}_{n}-\tilde{g}_{n-1})(s) \parallel ,
\parallel (\tilde{h}_{n}-\tilde{h}_{n-1})(s) \parallel ,
\parallel (\rho_{n+1}-\rho_{n})(s) \parallel  \leq \epsilon
\end{aligned}
\end{equation}
A) Estimation of $(\lambda_{n}')$, $(\mu_{n}')$,
$(\tilde{\mu}_{n}')$. We deduce from
(\ref{eq:4.37})-(\ref{eq:4.38}), taking $\partial = \partial_{r}$
that:
\begin{equation}\label{eq:4.42}
\partial R_{n}(s) = e^{(\mu_{n} - \lambda_{n})(s,r)}\xi_{n}(s)
\end{equation}
\begin{equation}\label{eq:4.43}
\partial W_{n}(s) = \eta_{n}(s)
  - (\sqrt{1+w^{2}+F/s^{2}}
\dot{\lambda}_{n})\xi_{n}(s)
\end{equation}
 We have,
using (\ref{eq:4.2}), (\ref{eq:4.42}), (\ref{eq:4.43})
\begin{equation}\label{eq:4.44}
\begin{aligned}
\parallel (\partial_{r} f_{n+1}-\partial_{r} f_{n})(s)\parallel & \leq
 \parallel \partial _{r,w}\overset{\circ}{f}\parallel
 (|\partial_{r} R_{n+1}-\partial_{r} R_{n}| +
|\partial_{r} W_{n+1}-\partial_{r} W_{n}
 |)(s)\\
 & \leq C\parallel \partial _{r,w}\overset{\circ}{f}\parallel
 (|e^{\mu_{n+1}-\lambda_{n+1}}\xi_{n+1}-
 e^{\mu_{n}-\lambda_{n}}\xi_{n}|\\
 & + |\eta_{n+1}-\eta_{n}| + |\dot{\lambda}_{n+1}\xi_{n+1} -
 \dot{\lambda}_{n}\xi_{n}|)(s)
\end{aligned}
\end{equation}
(\ref{eq:4.44}) gives, using (\ref{eq:4.41}) and since
$(\lambda_{n}),(\mu_{n}), (\xi_{n}), (\dot{\lambda_{n}})$ are
bounded,
\begin{equation}\label{eq:4.45}
\parallel (\partial_{r} f_{n+1}-\partial_{r} f_{n})(s)\parallel \leq
C(|\xi_{n+1}- \xi_{n}| + |\eta_{n+1}-\eta_{n}|)(s) + C\epsilon
\end{equation}
Next we have, using (\ref{eq:4.41}) and remark $4.2$ :
\begin{equation}\label{eq:4.46}
|(\tilde{g}_{n} \tilde{h}_{n})' - (\tilde{g}_{n-1}
\tilde{h}_{n-1})'|  \leq C \epsilon + C(|\tilde{g}_{n}'-
\tilde{g}_{n-1}'| + |\tilde{h}_{n}'- \tilde{h}_{n-1}'|)
\end{equation}
Now using (\ref{eq:4.41}) and the fact that $(\tilde{g}_{n})$,
$(\tilde{h}_{n})$, $(\tilde{g}_{n}')$, $\tilde{h}_{n}'$ are
bounded, we obtain
\begin{equation}\label{eq:4.47}
|\frac{1}{2}(\tilde{g}_{n}^{2} + \tilde{h}_{n}^{2})'-
\frac{1}{2}(\tilde{g}_{n-1}^{2} + \tilde{h}_{n-1}^{2})'|  \leq C
(|\tilde{g}_{n}' - \tilde{g}_{n-1}'| + |\tilde{h}_{n}' -
\tilde{h}_{n-1}'|) + C\epsilon
\end{equation}
We then deduce, from the expressions of $\rho_{n}$, $p_{n}$,
$j_{n}$ and using (\ref{eq:4.40}), (\ref{eq:4.45}),
(\ref{eq:4.46}), (\ref{eq:4.47}), (\ref{eq:4.41})
\begin{equation}\label{eq:4.48}
\parallel (\rho_{n+1}' - \rho_{n}')(s) \parallel, \parallel(
p_{n+1}'-p_{n}')(s)
\parallel, \parallel (j_{n+1}'-j_{n}')(s)
\parallel \leq C\epsilon + C(\gamma_{n} + \gamma_{n-1})(s)
\end{equation}
Concerning $(\mu_{n}')$, we obtain by taking the derivative of
(\ref{eq:4.4}) with respect to $r$, subtracting the relations
written for $n+1$ and $n$, after using (\ref{eq:4.41}),
(\ref{eq:4.48}) and the fact that $\mu_{n}$, $\mu_{n}'$ are
bounded :
\begin{equation}\label{eq:4.49}
\parallel (\mu_{n+1}' - \mu_{n}')(s)\parallel \leq C\epsilon +
C\int_{s}^{1}(\gamma_{n-1}(\tau) + \gamma_{n}(\tau)) d\tau
\end{equation}
Concerning $(\lambda_{n}')$, if we take the derivative of
(\ref{eq:4.5}) with respect to $r$, we have
\begin{equation}\label{eq:4.50}
\dot{\lambda}_{n}' = (8 \pi t\mu_{n}'\rho_{n} + 4 \pi t
\rho_{n}')e^{2\mu_{n}} - k \frac{\mu_{n}'}{t}e^{2\mu_{n}}
\end{equation}
We first deduce that $\dot{\lambda_{n}}'$ is bounded.  Next,
subtracting (\ref{eq:4.50}) written for $n+1$ and $n$, we obtain,
using (\ref{eq:4.40}), (\ref{eq:4.41}), (\ref{eq:4.48}), and since
$(\mu_{n})$, $(\rho_{n})$, $(\mu_{n}')$ are bounded
\begin{equation}\label{eq:4.51}
\parallel (\dot{\lambda}_{n+1}' - \dot{\lambda}_{n}')(s) \parallel
\leq C\epsilon + C(\gamma_{n-1}(s) + \gamma_{n}(s))
\end{equation}
Now, from (\ref{eq:4.6}), using (\ref{eq:4.51}) and integrating
over $[s,1]$, we obtain :
\begin{equation}\label{eq:4.52}
\parallel (\lambda_{n+1}' - \lambda_{n}')(s) \parallel
\leq C \epsilon + C \int_{s}^{1}(\gamma_{n-1}(\tau) +
\gamma_{n}(\tau)) d \tau
\end{equation}
We will also need to bound $\tilde{\mu}_{n+1}' -
\tilde{\mu}_{n}'$. If we take the derivative of (\ref{eq:4.7})
with respect to $r$, we obtain, after subtracting the expressions
written for $n+1$ and $n$ and using (\ref{eq:4.40}),
(\ref{eq:4.41}), (\ref{eq:4.48}):
\begin{equation}\label{eq:4.53}
\parallel (\tilde{\mu}_{n+1}' - \tilde{\mu}_{n}')(s) \parallel
\leq C \epsilon + C (\gamma_{n-1}(s) + \gamma_{n}(s))
\end{equation}
B) Estimation of $(\tilde{g}_{n}')$, $(\tilde{h}_{n}')$. Recall
that
\begin{equation}\label{eq:4.54}
\begin{aligned}
X_{n} = \tilde{g}_{n} - \tilde{h}_{n} \ ; \ Y_{n} = \tilde{g}_{n}
+ \tilde{h}_{n} \ ;\ a_{n} =
(-\dot{\lambda}_{n}-\frac{1}{t})e^{-\mu_{n}}-\tilde{\mu}_{n}e^{-\lambda_{n}};
\\  b_{n} = -\frac{e^{-\mu_{n}}}{t} \ ; \ c_{n} =
(-\dot{\lambda}_{n}-\frac{1}{t})e^{-\mu_{n}}+
\tilde{\mu}_{n}e^{-\lambda_{n}}
\end{aligned}
\end{equation}
Set
\begin{equation}\label{eq:4.55}
\begin{aligned}
C_{n} = a_{n}X_{n}+ b_{n}Y_{n} \ ; \ \  \ \tilde{C}_{n}=
(\lambda_{n}'-\mu_{n}')e^{-\lambda_{n}}X_{n+1}' + \mu_{n}'C_{n} +
C_{n}'\\ D_{n} = b_{n}X_{n}+ c_{n}Y_{n} \ ; \ \  \ \tilde{D}_{n}=
(\mu_{n}'-\lambda_{n}')e^{-\lambda_{n}}Y_{n+1}' + \mu_{n}'D_{n} +
D_{n}'
\end{aligned}
\end{equation}
Then system (\ref{eq:4.8})-(\ref{eq:4.9}) becomes
\begin{equation*}
\begin{cases}
D_{n}^{+}X_{n+1} = C_{n} \\
 D_{n}^{-}Y_{n+1} = D_{n}
\end{cases}
\end{equation*}
If we take the derivative of the above system with respect to $r$,
a direct calculation shows that
\begin{equation}\label{eq:4.56}
D_{n}^{+}X_{n+1}' = \tilde{C}_{n}
\end{equation}
\begin{equation}\label{eq:4.57}
D_{n}^{-}Y_{n+1}' = \tilde{D}_{n}
\end{equation}
Consider the following characteristic curves $\gamma_{n}^{1}$,
$\gamma_{n}^{2}$ of the wave operator, starting from the point
$(s,r)$, i.e for every $n$,
\begin{equation}\label{eq:4.58}
\dot{\gamma}_{n}^{1}= e^{\lambda_{n}-\mu_{n}} \ ,\dot{
\gamma}_{n}^{2}= -e^{\lambda_{n}-\mu_{n}} \ ,
\gamma_{n}^{1}(s)=\gamma_{n}^{2}(s)=r
\end{equation}
We have $D_{n}^{+} = 2e^{-\mu_{n}}\frac{d}{dt}$ on
$\gamma_{n}^{1}$ and $D_{n}^{-} = 2e^{-\mu_{n}}\frac{d}{dt}$ on
$\gamma_{n}^{2}$. We then have, integrating (\ref{eq:4.56}) over
$\gamma_{n}^{1}$ and (\ref{eq:4.57}) over $\gamma_{n}^{2}$, and
after subtraction,
\begin{equation}\label{eq:4.59}
(X_{n+1}'-X_{n}')(s) =
\frac{1}{2}\int_{s}^{1}[e^{\mu_{n-1}}\tilde{C}_{n-1}(\tau,\gamma_{n-1}^{1}(\tau))
-e^{\mu_{n}}\tilde{C}_{n}(\tau,\gamma_{n}^{1}(\tau))]d\tau
\end{equation}
\begin{equation}\label{eq:4.60}
(Y_{n+1}'-Y_{n}')(s) =
\frac{1}{2}\int_{s}^{1}[e^{\mu_{n-1}}\tilde{D}_{n-1}(\tau,\gamma_{n-1}^{2}(\tau))
-e^{\mu_{n}}\tilde{D}_{n}(\tau,\gamma_{n}^{2}(\tau))]d\tau
\end{equation}
Since $e^{\mu_{n}}$ and $\tilde{C}_{n}$ are bounded, we have
\begin{equation}\label{eq:4.61}
\begin{aligned}
|e^{\mu_{n}}\tilde{C}_{n}(\tau,\gamma_{n}^{1}(\tau)) -
e^{\mu_{n-1}}\tilde{C}_{n-1}(\tau,\gamma_{n-1}^{1}(\tau))|  \leq
C[\parallel (e^{\mu_{n}}-e^{\mu_{n-1}})(\tau)\parallel \\ +
\parallel (\tilde{C}_{n}-\tilde{C}_{n-1})(\tau)\parallel
+ |\tilde{C}_{n-1}(\tau,\gamma_{n}^{1}(\tau))
-\tilde{C}_{n-1}(\tau,\gamma_{n-1}^{1}(\tau))|
\\  +|e^{\mu_{n-1}}(\tau,\gamma_{n}^{1}(\tau)) -
e^{\mu_{n-1}}(\tau,\gamma_{n-1}^{1}(\tau))|]
\end{aligned}
\end{equation}
Now integrating the relation
$\dot{\gamma}_{n}^{1}-\dot{\gamma}_{n-1}^{1} =
e^{\lambda_{n}-\mu_{n}}-e^{\lambda_{n-1}-\mu_{n-1}}$ over
$[s,\tau]$ yields
\begin{equation}\label{eq:4.62}
|\gamma_{n}^{1}-\gamma_{n-1}^{1}|(\tau) \leq C \sup\{
\parallel (\lambda_{n}-\lambda_{n-1})(t)\parallel + \parallel (\mu_{n}-\mu_{n-1})(t) \parallel,
T_{4}\leq t\leq 1\}
\end{equation}
we then deduce from (\ref{eq:4.36}), (\ref{eq:4.62}),
 the uniform continuity of $(e^{\mu_{n-1}})$,
$(\tilde{C}_{n-1})$ over the compact set $K = [T_{4},1]\times
(\gamma_{n}^{1}([T_{4},1])\cup \gamma_{n-1}^{1}([T_{4},1]))$,
(\ref{eq:4.61}) and (\ref{eq:4.59})-(\ref{eq:4.60}), that
\begin{equation}\label{eq:4.63}
|X_{n+1}'-X_{n}'|(s) \leq C\epsilon + C\int_{s}^{1}\parallel
(\tilde{C}_{n} -\tilde{C}_{n-1})(\tau)\parallel d\tau
\end{equation}
\begin{equation}\label{eq:4.64}
|Y_{n+1}'-Y_{n}'|(s) \leq C\epsilon + C\int_{s}^{1}\parallel
(\tilde{D}_{n} -\tilde{D}_{n-1})(\tau)\parallel d\tau
\end{equation}
Therefore, for $n$ sufficiently large,
\begin{equation}\label{eq:4.65}
\begin{aligned}
\parallel (\tilde{g}_{n+1}'-\tilde{g}_{n}')(s)\parallel,
&\parallel (\tilde{h}_{n+1}'-\tilde{h}_{n}')(s)\parallel   \leq
C\epsilon + C\int_{s}^{1}[\parallel (\tilde{C}_{n}
-\tilde{C}_{n-1})(\tau)\parallel \\ &+  \parallel (\tilde{D}_{n}
-\tilde{D}_{n-1})(\tau)\parallel] d\tau
\end{aligned}
\end{equation}
Now from (\ref{eq:4.55}), (\ref{eq:4.41}) and the fact that the
sequences $(\dot{\lambda}_{n})$, $(\lambda_{n})$, $(\mu_{n})$,
$(\tilde{\mu}_{n}')$ are bounded together with their first derivatives,
we have
\begin{equation*}
\parallel (\tilde{C}_{n} - \tilde{C}_{n-1})(\tau) \parallel
\leq C \epsilon + C (\gamma_{n-2}(\tau)+ \gamma_{n-1}(\tau) +
\gamma_{n}(\tau))
\end{equation*}
and
\begin{equation*}
\parallel( \tilde{D}_{n} - \tilde{D}_{n-1})(\tau) \parallel
\leq C \epsilon + C (\gamma_{n-2}(\tau)+ \gamma_{n-1}(\tau) +
\gamma_{n}(\tau))
\end{equation*}
Therefore, we deduce from (\ref{eq:4.65}) that
\begin{equation}\label{eq:4.66}
\parallel (\tilde{g}_{n+1}'-\tilde{g}_{n}')(s)\parallel \ ,
\parallel (\tilde{h}_{n+1}'-\tilde{h}_{n}')(s)\parallel \leq C\epsilon +
C\int_{s}^{1}(\gamma_{n-2}(\tau)+ \gamma_{n-1}(\tau) +
\gamma_{n}(\tau)) d\tau
\end{equation}
C) Convergence of $(\lambda_{n}')$, $(\mu_{n}')$,
$(\tilde{\mu}_{n}')$, $(\tilde{g}_{n}')$, $(\tilde{h}_{n}')$. Add
inequalities
(\ref{eq:4.39}), (\ref{eq:4.49}),(\ref{eq:4.52}) and (\ref{eq:4.66})
and take the supremum over $s \in [t,1]$ to obtain using
(\ref{eq:4.40})
\begin{equation}\label{eq:4.67}
 \gamma_{n}(t) \leq C\epsilon +
C\int_{t}^{1}(\gamma_{n-2}(s)+ \gamma_{n-1}(s) + \gamma_{n}(s)) ds
\end{equation}
Define $\Gamma_{n}(t) = \sup \{\gamma_{m},\ m \leq n \}$; then
$(\Gamma_{n})$ is an increasing sequence and (\ref{eq:4.67}) gives
\begin{equation*}
 \Gamma_{n}(t) \leq C\epsilon +
C\int_{t}^{1} \Gamma_{n}(s) ds
\end{equation*}
And by Gronwall's lemma,
\begin{equation*}
 \Gamma_{n}(t) \leq C\epsilon ,\ \ n > N, \ \ t\in [T_{4},1]
\end{equation*}
We then deduce that $(\Gamma_{n})$ converges uniformly to $0$, and
from (\ref{eq:4.40})-(\ref{eq:4.53}), $(\lambda_{n}')$,
$(\mu_{n}')$, $(\tilde{g}_{n}')$, $(\tilde{h}_{n}')$,
$(\tilde{\mu}_{n}')$ converge uniformly on $[T_4,1]$. We deduce
from system (\ref{eq:4.8})-(\ref{eq:4.9}), the uniform convergence
of $(\dot{\tilde{g}}_{n})$ and $(\dot{\tilde{h}}_{n})$. The
regularity of $f$, $\lambda$, $\mu$, $\tilde{g}$, $\tilde{h}$ (and
$\tilde{\mu}$) follows from step $1$ and step $2$. Therefore $g =
e^{\mu}\tilde{g}$ and $h = e^{\lambda}\tilde{h}$ are regular. Note
that, using the convergence of the derivatives, we can prove that
the limit $(f, \lambda, \mu, \tilde{\mu}, g, h)$ is a regular
solution of (\ref{eq:2.3}), (\ref{eq:1.3}), (\ref{eq:1.4}),
(\ref{eq:2.4}), (\ref{eq:2.5}), (\ref{eq:2.6}) and by Proposition
\ref{P:3.2}, we conclude the existence of a regular function
$\phi$ such that $(f, \lambda, \mu, \phi)$ is a solution of the
full system (\ref{eq:1.2})-(\ref{eq:1.11}). To end this theorem,
we prove the uniqueness of the solution.
 Let  $u_{i} = (f_{i},
\lambda_{i}, \mu_{i}, \phi_{i})$, $i = 1,2$  be two regular
solutions of the Cauchy problem for the same initial data
$(\overset{\circ}{f}, \overset{\circ}{\lambda},
\overset{\circ}{\mu}, \overset{\circ}{\phi}, \psi) $ at $t = 1$.
Using the fact that $u_{i}$ is a solution of the system, one
proceeds similarly as to prove the convergence of iterates to
obtain
\begin{equation*}
\alpha(t) \leq C\int_{t}^{1}\alpha(s)ds
\end{equation*}
where
\begin{align*}
\alpha(t) &= \sup \{ \parallel f_{1}(s) - f_{2}(s)\parallel +
\parallel \lambda_{1}(s) - \lambda_{2}(s)\parallel + \parallel \mu_{1}(s) -
\mu_{2}(s)\parallel\\
 & + \parallel \tilde{g}_{1}(s) - \tilde{g}_{2}(s)\parallel +
 \parallel \tilde{h}_{1}(s) - \tilde{h}_{2}(s)\parallel \ ; \ s \in [t, 1]
 \},
\end{align*}
with $\tilde{g}_{1} = \dot{\phi}_{1}e^{\mu_{1}}$; \ $\tilde{g}_{2}
= \dot{\phi}_{2}e^{\mu_{2}}$; \ $\tilde{h}_{1} =
\phi_{1}'e^{\lambda_{1}}$; \ $\tilde{h}_{2} =
\phi_{2}'e^{\lambda_{2}}$. We deduce that $\alpha(t) = 0$, \ for
$t \in ]0,1]$. This implies that $f_{1}=f_{2}$, $\lambda_{1}=
\lambda_{2}$, $\mu_{1}= \mu_{2}$, $\tilde{g}_{1}= \tilde{g}_{2}$
and $\tilde{h}_{1}= \tilde{h}_{2}$. By Proposition \ref{P:3.2}, we
conclude that $\phi_{1}= \phi_{2}$.$\square$

    Let us now prove continuation criteria for $t$ decreasing.
\begin{theorem} \label{T:4.5}
Let $(\overset{\circ}{f}, \overset{\circ}{\lambda},
\overset{\circ}{\mu}, \overset{\circ}{\phi}, \psi)$
be initial data
as in theorem $4.4$. If $k = -1$, assume that $\overset{\circ}{\mu} <0$.
Let $(f, \lambda, \mu, \phi)$ be a
solution of (\ref{eq:1.2})-(\ref{eq:1.11}) on a left maximal
interval of existence $]T, 1]$ for which
\begin{eqnarray*}
\sup \{ |w| | (t, r, w, F) \in {\rm supp} f \} < \infty
\end{eqnarray*}

\noindent
 If $\sup\{(e^{-2\mu}\dot\phi^2+e^{-2\lambda}\phi'^2)(t,r);t\in ]T,1]\}
<\infty$ then $T=0$.
\end{theorem}
{\bf Proof} : Let $(f, \lambda, \mu, \tilde{\mu}, g, h)$ be a left
maximal solution of the auxiliary system (\ref{eq:2.3}),
(\ref{eq:1.3}), (\ref{eq:1.4}), (\ref{eq:2.4}), (\ref{eq:2.5}),
(\ref{eq:2.6}) with existence interval $]T, 1]$. By Proposition
\ref{P:3.2}, there exists $\phi$ such that  $(f, \lambda, \mu,
\phi)$ solves (\ref{eq:1.2})-(\ref{eq:1.11}). By assumption
\begin{eqnarray*}
Q^{\ast} := \sup \{  |w| | (r, w, F) \in {\rm supp} f(t), \ t \in
]T, 1] \} < \infty.
\end{eqnarray*}
 We want to show that $T = 0$, so let us assume that $T > 0$
and take $t_{1} \in ]T, 1[$. We will show that the system has a
solution with initial data $(f(t_{1}), \lambda(t_{1}), \mu(t_{1}),
\phi(t_{1}),\\ \dot{\phi}(t_{1}))$ prescribed at $t = t_{1}$ which
exists on an interval $[t_{1} - \delta, t_{1}]$ with $\delta > 0$
independent of $t_{1}$. By moving $t_{1}$ close enough to $T$ this
would extend our initial solution beyond $T$, a contradiction to
the initial solution being left maximal. We have proved in
Proposition \ref{P:4.1} that such a solution exists at least on
the left maximal existence interval of the solutions $(z_{1},
z_{2})$ of
\begin{eqnarray*}
z_{1}(t) = W(t_{1}) + K(t_{1}) +
C_{1}\int_{t}^{t_{1}}\frac{1}{s}(1+z_{1}(s))^{5} ds
\end{eqnarray*}
\begin{eqnarray*}
z_{2}(t) = C_{4} \exp\left[\int_{t}^{t_{1}}C_{2}(s) (C_{3} +
z_{2}(s)) ds\right],
\end{eqnarray*}
where
\begin{eqnarray*}
W(t_{1}) := \sup \{  |w| | (r, w, F) \in {\rm supp} f(t_{1}) \},
\end{eqnarray*}
\begin{eqnarray*}
K(t_{1}) := \sup \{( |\dot{\phi}^{2}|e^{-2\mu} +
|{\phi'}^{2}|e^{-2\lambda})^{\frac{1}{2}}(t_{1},r) \ ; \ r \in
\mathbb{R}\},
\end{eqnarray*}
\begin{eqnarray*}
A(t_{1}) := \sup \{[(\dot{\phi}' - \mu'\dot{\phi})^{2}e^{-2\mu} +
(\phi'' - \lambda'\phi')^{2}e^{-2\lambda}](t_{1},r) \ ; \ r \in
\mathbb{R}\},
\end{eqnarray*}
\begin{eqnarray*}
C_{1} =  \frac{C}{C_0}(1+F_{0})^{2}(1+ \parallel f(t_1)\parallel)
\ ; \ \ C_{0} =\inf\{e^{-\mu(t_{1},r)}; \ r \in \mathbb{R}\}
\end{eqnarray*}
\begin{eqnarray*}
C_3 := \parallel e^{-2\mu(t_1)} \mu'(t_1)\parallel + \parallel
\lambda'(t_1)\parallel + 1 \ ; \ \
 C_4 := A(t_{1})+ \parallel
\partial_{(r,w)}f(t_{1}) \parallel
\end{eqnarray*}
and $C_2$ is an increasing function of $z_1$. Now $W(t_1) \leq
Q^{\ast}$, $\parallel f(t_1)\parallel = \parallel
\overset{\circ}{f}\parallel$, $F_0$ is unchanged since $F$ is
constant along characteristics, and since $t_1 < 1$, taking
$t=t_1$ in (\ref{eq:4.4}) shows that: $C_0(\mu(t_1)) \geq
C_{0}(\overset{\circ}{\mu})$. Thus there exists a constant $M_{1}
> 0$ such that
\begin{eqnarray*}
\frac{1}{T}C_{1}(f(t_1), F_0, \mu(t_1)) \leq M_{1}
  \ \ for \ \ t_{1}  \in ]T, 1].
\end{eqnarray*}
The expressions of $\rho, p, j, \dot{\lambda}, \tilde{\mu}$ show
since $|w| \leq Q^{*}$, that
\begin{equation*}
|\rho(s,r)|, \ |p(s,r)|, \ |j(s,r)|,\ |\dot{\lambda}(s,r)|, \
|\tilde{\mu}e^{\mu-\lambda}(s,r)| \leq C+(K(s))^{2}
\end{equation*}
 Since $K(t)$ is
bounded on $]T, 1]$, we proceed as in step $6$ of the proof of
theorem $3.1$ in \cite{rein1} to prove that $\partial_{(r,w)}f$ is
uniformly bounded on $]T,1]$. Let $\mu' = \tilde{\mu}$; consider
the relations
\begin{equation}\label{eq:4.68}
\mu' = -4 \pi te^{\lambda+\mu}j \ ; \ \ \mu'' = -[(\lambda'+\mu')j
+ j']4\pi te^{\lambda+\mu}
\end{equation}
\begin{equation}\label{eq:4.69}
\dot{\lambda}' = e^{2\mu}(8 \pi t\mu'\rho + \frac{k\mu'}{t} + 4\pi
t\rho') \ ; \ \lambda' = \overset{\circ}{\lambda}' +
\int_{1}^{t}\dot{\lambda}'(s,r)ds
\end{equation}
We bound $\rho'$ , $j'$ by  quantities which depend on
$\partial_{r}f$ and $A(s)$. Since $\partial_{r}f$, $\rho$ and $j$
are bounded, we deduce from the above relations that $\mu''$,
$\dot{\lambda}'$ and consequently $v(s)$ and $h(s)$ (see
Proposition $2.5$) are bounded each by $A(s)$. Using inequality
(\ref{eq:2.10}) and the fact that $K(s)$ is bounded, we obtain
\begin{equation*}
A(t) \leq A_{0} + C \int_{t}^{1}(A(s)+1)ds
\end{equation*}
And we deduce that $A(t) \leq A^{*} = (1+A_{0})e^{C}$; $t \in
]T,1]$.
 Therefore
\begin{eqnarray*}
D := \sup \{ \parallel \partial_{(r, w)}f(t)\parallel + A(t) \  |
t \in ]T, 1] \} < \infty.
\end{eqnarray*}
{}From
\begin{eqnarray*}
\mu'(t, r) & = \frac{e^{2\mu}}{t}\left( \overset{\circ}{\mu}'(r)
e^{-2 \overset{\circ}{\mu}} + 4 \pi \int_{1}^{t}p'(s, r) s^{2}
ds\right),
\end{eqnarray*}
(\ref{eq:4.68}), (\ref{eq:4.69}) and since $K(t)$ and $A(t)$ are
bounded, $p'$ is bounded and we obtain a uniform bound
$C_{3}(\mu(t_1), \lambda(t_1)) \leq M_{3}$. Let $y_{2}$ be the
left maximal solution of
\begin{eqnarray*}
y_{2}(t) = D \exp\left[\int_{t}^{t_1}C^{*}_{2}(s)(M_{3} +
y_{2}(s)) ds \right],
\end{eqnarray*}
where $C^{*}_{2}$ depends on $y_{1}$ in the same way as $C_{1}$
depends on $z_{1}$.  $y_{2}$ exists on an interval $[t_{1} -
\alpha, t_{1}]$ with $\alpha > 0$ independent of $t_1$. If we
choose $t_{1}$ such that $T > t_{1}-\alpha$ then $z_{1}$ is
bounded $z_{2} \leq y_{2}$ by construction. In particular, $z_{2}$
exists on $I \subset [t_{1} - \alpha, t_{1}]$, and the proof of the theorem is complete.$\square$

We prove in the next theorem, the analogue of theorems $4.4$ and
$4.5$ for $t \geq 1$.
\begin{theorem}\label{T:4.7}
Let $(\overset{\circ}{f}, \overset{\circ}{\lambda},
\overset{\circ}{\mu}, \overset{\circ}{\phi}, \psi)$ be initial
data as in Theorem \ref{T:4.4}. Then there exists a unique, right
maximal, regular solution $(f, \lambda, \mu, \phi)$ of
(\ref{eq:1.2})-(\ref{eq:1.11}) with $(f, \lambda, \mu, \phi,
\dot{\phi} )(1) = (\overset{\circ}{f}, \overset{\circ}{\lambda},
\overset{\circ}{\mu}, \overset{\circ}{\phi}, \psi)$ on a time
interval $[1, T[$ with $T \in ]1, \infty]$. If
\begin{eqnarray*}
\sup \{ \mid w \mid | (t, r, w, F) \in {\rm supp} f \} < \infty ;
\end{eqnarray*}
\begin{eqnarray*}
\sup \{ e^{2\mu(t, r)} | r \in \mathbb{R}, \ t\in [1, T[ \} <
\infty
\end{eqnarray*}
and
\begin{eqnarray*}
  K(t) < \infty
\end{eqnarray*}
 then $T = \infty$.
\end{theorem}
{\bf Proof} : We give only those parts of the proof which differ
from the proof of Theorem \ref{T:4.4} for $t \leq 1$. The iterates
are defined in the same way as before, except that now
(\ref{eq:4.4}) is used to define $\mu_{n}$ only on the interval
$[1, T_{n}[$, where
\begin{eqnarray*}
 T_{n}  :=  \sup \left\{ \tau \in ]1, T_{n-1}[ |
\frac{e^{-2 \overset{\circ}{\mu}(r)}}{t} - \frac{8
\pi}{t}\int_{1}^{t}s^{2}p_{n}(s, r) ds > 0, r \in \mathbb{R}, t
\in [1, \tau]\right\},
\end{eqnarray*}
$[1, T_{n-1}[$ being the existence interval of the previous
iterates and $T_0 = \infty$. Define:
\begin{displaymath}
Q_{n}(t) := \sup \left\{ |w|, \ (r,w,F) \in \supp f_{n}(t)
\right\} \ , \ t \in [1,T_{n}[
\end{displaymath}
\begin{displaymath}
E_{n}(t) := \sup \left\{ se^{2\mu_{n}(s, r)} | r \in \mathbb{R}, 1
\leq s \leq t \right\}
\end{displaymath}
we obtain the estimates
\begin{equation*}
\sqrt{1+w^{2}+F/{t^{2}}} \leq \sqrt{1+(Q_{n}(t))^{2}+F_{0}} \leq
(1+F_{0})(1+Q_{n}(t));
\end{equation*}
\begin{equation}\label{eq:4.70}
\parallel \rho_{n}(t)\parallel, \parallel p_{n}(t)\parallel,
 \parallel j_{n}(t)\parallel \leq \frac{C^{*}}{t}(1+Q_{n}(t))^{2} +
 (K_{n-1}(t))^{2};
\end{equation}
 and
\begin{displaymath}
\mid e^{\mu_{n}-\lambda_{n}} \tilde{\mu}_{n}(t,r)\mid + \mid
\dot{\lambda}_{n}(t,r)\mid \leq C^{*}(1+Q_{n}(t))^{2}
 +(1+K_{n-1}(t))^{2}(1+E_{n}(t)).
\end{displaymath}
where $C^{*} = C(1+F_{0})^{2}
 (1+\parallel \overset{\circ}{f}\parallel)$.
Thus, we have similarly to (\ref{eq:4.16}):
\begin{equation}\label{eq:4.71}
Q_{n+1}(t) \leq W_{0} + C^{*}
 \int_{1}^{t}(1+Q_{n}(s))^{2}(1+E_{n}(s))(1+K_{n-1}(s))^{2}(1+Q_{n+1}(s))ds.
\end{equation}
and similarly to (\ref{eq:4.18}):
\begin{equation}\label{eq:4.72}
K_{n+1}(t) \leq K_{0} + C^{*}
 \int_{1}^{t}(1+Q_{n}(s))^{2}(1+K_{n-1}(s))^{2}K_{n}(s)(1+E_{n}(s))ds.
\end{equation}
We deduce from the field equation (\ref{eq:1.4}) that
\begin{eqnarray*}
(2\dot{\mu}_{n}e^{2\mu_{n}})t = e^{2\mu_{n}} + ke^{4\mu_{n}} +
8\pi (t e^{2\mu_{n}})^{2} p_{n}
\end{eqnarray*}
Integrating over $[1, s]$ and using integration by parts for the
left hand side yields the following estimate, since
(\ref{eq:4.70}) holds,
\begin{equation}\label{eq:4.73}
E_{n}(t) \leq \parallel e^{2 \overset{\circ}{\mu}}\parallel +
C^{*}
 \int_{1}^{t}(1+Q_{n}(s))^{2}(1+E_{n}(s))^{2}(1+K_{n-1}(s))^{2} ds.
\end{equation}
Reasoning in the same way as in the proof of Proposition
\ref{P:4.1}, we can say the differential inequalities
(\ref{eq:4.71}), (\ref{eq:4.72}), (\ref{eq:4.73}) allow us to
estimate $Q_{n}$, $K_{n}$ and $E_{n}$ against the solution
$z_{1}$, $z_{2}$ and $z_{3}$ of the system
\begin{equation*}
z_{1}(t) = W_{0} + C^{*}
 \int_{1}^{t}(1+z_{1}(s))^{3}(1+z_{2}(s))^{2}(1+z_{3}(s))
 ds,
\end{equation*}
\begin{equation*}
z_{2}(t) = K_{0} + C^{*}
 \int_{1}^{t}(1+z_{1}(s))^{2}(1+z_{2}(s))^{3}](1+z_{3}(s))ds,
\end{equation*}
\begin{equation*}
z_{3}(t) = \parallel e^{2 \overset{\circ}{\mu}}\parallel + C^{*}
 \int_{1}^{t}(1+z_{1}(s))^{2}(1+z_{2}(s))^{2}(1+z_{3}(s))^{2}
 ds,
\end{equation*}
and in particular $T_{n} \geq T$ where $[1, T[$ is the right
maximal existence interval of $(z_{1}, z_{2}, z_{3})$.
 One can now establish a bound on first order derivatives of the iterates
in the same way as in the proof of Proposition \ref{P:4.1} and
obtains a local solution on a right maximal existence interval
which is extendible if the quantities $Q(t)$ , $E(t) =
\parallel e^{2\mu(t)}\parallel$ and $K(t)$ can be
bounded.$\square$
\begin{theorem}\label{T:4.8}
 Let $(f, \lambda, \mu, \phi)$ be a right maximal regular solution
 obtained in Theorem \ref{T:4.7}. Assume that
 \begin{eqnarray*}
\sup \{ \mid w \mid | (t, r, w, F) \in {\rm supp} f \} < \infty ;
\end{eqnarray*}
and
\begin{eqnarray*}
\sup \{ e^{2\mu(t, r)} | r \in \mathbb{R}, \ t\in [1, T[ \} < C <
\infty ;
\end{eqnarray*}
then \ \quad \   $T = \infty$.
\end{theorem}
{\bf Proof} : We deduce from system (\ref{eq:2.1})-(\ref{eq:2.2}):
\begin{equation*}
D^{+}X^2 = 2aX^2 + 2bXY
\end{equation*}
\begin{equation*}
 D^{-}Y^2 = 2bXY + 2cY^2
\end{equation*}
On the characteristic curves of the wave equation, we obtain
\begin{equation}\label{eq:4.74}
\frac{d}{dt}X^2(t,\gamma_{1}(t)) =  2e^{\mu}(aX^2 + b
XY)(t,\gamma_{1}(t))
\end{equation}
\begin{equation}\label{eq:4.75}
 \frac{d}{dt}Y^2(t,\gamma_{2}(t)) =  2e^{\mu}(bXY
+ cY^2)(t,\gamma_{2}(t))
\end{equation}
From (\ref{eq:4.74}), we have:
\begin{align*}
\frac{d}{dt}X^2(t,\gamma_{1}(t))
&=2(-\dot{\lambda}-\mu'e^{\mu-\lambda}-\frac{1}{t})X^2 -
\frac{2XY}{t}\\
&=8\pi te^{2\mu}(j-\rho)X^2 + \frac{1+ke^{2\mu}}{t}X^2 -\frac{2}{t}X^2
- \frac{2XY}{t}\\
&\leq (\frac{-1}{t}+ \frac{ke^{2\mu}}{t})X^2 +
\frac{X^2+Y^2}{t} \ \ \ \
{\rm since} \ \ \ j-\rho < 0 \ ; \\
& \leq (\frac{Ck}{t}X^2 + \frac{1}{t}Y^2)(t,\gamma_{1}(t))
\end{align*}
 Since $j+\rho >0$, we deduce as above, from (\ref{eq:4.75}), the estimate
\begin{align*}
\frac{d}{dt}Y^2(t,\gamma_{2}(t)) \leq (\frac{1}{t}X^2 +
\frac{Ck}{t}Y^2)(t,\gamma_{2}(t))
\end{align*}
After integrating these two inequalities over $[1,t]$ and taking
the maximum over space, we obtain :
\begin{equation*}
K(t)^2 \leq K(1)^2 + \int_1^t (\frac{1+Ck}{s})K(s)^2ds
 \end{equation*}
We deduce by Gronwall's lemma that:
\begin{equation}\label{eq:4.76}
K(t) \leq K(1)t^\frac{1+Ck}{2} \ , \ \ \ for \  t \  \in \ [1, T[
\end{equation}
And we conclude by Theorem \ref{T:4.7} that $T= \infty$. $\square$
\section{Future global existence in a particular case}
 In the case $k=1$, there is no global existence in the future
since this already fails without a scalar field, as shown in
\cite{rein1} and \cite{rein}.
 Now we prove in the special
case where there is only a scalar field (i.e. $f$ is identically
zero), the existence of the solution for all $t \geq 1$ for $k=0$
or $k=-1$.
\begin{theorem}\label{T:5.1}
 Suppose that $f=0$ and let $(\lambda, \mu, \phi)$ be a right maximal regular
solution
 obtained in Theorem \ref{T:4.7}. $T = \infty$ for $k=0$ or $k=-1$.
\end{theorem}
{\bf Proof} : we establish a series of estimates which will result
in an upper bound on $\mu$ and will therefore prove that $T =
\infty$. Similar estimates were used in \cite{andreasson}. Unless
otherwise specified in what follows constants denoted by $C$ will
be positive, may depend on the initial data and may change their
value from line to line.
Firstly, integration of (\ref{eq:3.7}) (where $\bar{p}$ is replaced
by $p$) with respect to $t$ and the fact that $p$ is non-negative
imply that
\begin{equation}\label{eq:5.1}
e^{2 \mu(t, r)} = \left[\frac{(e^{-2 \overset{\circ}{\mu}(r)} +
k)}{t} - k - \frac{8 \pi}{t}\int_{1}^{t}s^{2}p(s, r)
ds\right]^{-1} \ \geq \frac{t}{C-kt}, \ t \in [1, T[
\end{equation}
Next let
 us prove that
\begin{equation}\label{eq:5.2}
\int_{0}^{1} e^{\mu+\lambda}\rho(t,r) dr \leq C t, \ t \in [1, T[
\end{equation}
A calculation shows that, since $\mu$ and $j$ are periodic
with respect to $r$ and $\rho=p$ :
\begin{equation*}
\frac{d}{dt}\int_{0}^{1} e^{\mu+\lambda}\rho(t,r) dr =
-\frac{1}{t}\int_{0}^{1}e^{\mu+\lambda}\left[2\rho+q
-\rho(1+ke^{2\mu})\right]dr
\end{equation*}
Now $q \geq -2\rho$ and $q+\rho \geq -\rho$ so that for $k=0$,
\begin{equation*}
\frac{d}{dt}\int_{0}^{1} e^{\mu+\lambda}\rho(t,r) dr \leq
\frac{1}{t}\int_{0}^{1} e^{\mu+\lambda}\rho(t,r) dr
\end{equation*}
and integrating this inequality with respect to $t$ yields
(\ref{eq:5.2}) for $k=0$. For $k=-1$, we have, using
(\ref{eq:5.1}):
\begin{align*}
 \frac{d}{dt}\int_{0}^{1} e^{\mu+\lambda}\rho(t,r) dr &=
  -\frac{1}{t}\int_{0}^{1}e^{\mu+\lambda}\left[2\rho+q
-\rho(1-e^{2\mu})\right]dr\\
& \leq -\frac{1}{t}\int_{0}^{1}e^{\mu+\lambda}(\rho+q)dr
-\frac{1}{C+t}\int_{0}^{1}e^{\mu+\lambda}
\rho dr\\
& \leq (\frac{1}{t}-\frac{1}{C+t})\int_{0}^{1}e^{\mu+\lambda}\rho
dr\\
& \leq \frac{1}{t}\int_{0}^{1}e^{\mu+\lambda}\rho dr
\end{align*}
 Integrating this
inequality with respect to $t$ yields (\ref{eq:5.2}) for $k=-1$.
 Using
(\ref{eq:1.5}), the fact that $|j| \leq \rho$ and (\ref{eq:5.2})
we find
\begin{align*}
\mid \mu(t,r)- \int_{0}^{1}\mu(t,\sigma)d\sigma \mid & =
 \mid \int_{0}^{1}\int_{\sigma}^{r}\mu'(t,\tau)d\tau d\sigma \mid
 \leq \int_{0}^{1}\int_{0}^{1}|\mu'(t,\tau)|d\tau d\sigma \\
 & \leq 4\pi t\int_{0}^{1}e^{\mu+\lambda}|j(t,\tau)|d\tau
 \leq 4\pi t\int_{0}^{1}e^{\mu+\lambda}\rho(t,\tau)d\tau
\end{align*}
that is
\begin{equation}\label{eq:5.3}
  \mid \mu(t,r)- \int_{0}^{1}\mu(t,\sigma)d\sigma \mid
  \leq Ct^{2}, \ t \in [1, T[, \ r \in [0,1]
\end{equation}
 Next we show that
\begin{equation}\label{eq:5.4}
  e^{\mu(t,r)-\lambda(t,r)} \leq Ct , \ t \in [1, T[, \ r
  \in [0,1].
\end{equation}
Using relations $\dot{\mu}-\dot{\lambda}= 4\pi
te^{2\mu}(p-\rho)+\frac{1+ke^{2\mu}}{t}$,\ \  $p-\rho = 0$ and
(\ref{eq:5.1}), we obtain
\begin{align*}
 \frac{ \partial}{\partial t}e^{\mu-\lambda} & = (\dot{\mu}-\dot{\lambda})e^{\mu-\lambda} =
 e^{\mu-\lambda}\frac{1+ke^{2\mu}}{t}\\
 & \leq (\frac{1}{t}+\frac{k}{C-kt})e^{\mu-\lambda};
\end{align*}
 and integrating this inequality with
respect to $t$ yields (\ref{eq:5.4}).

 We now estimate the average of $\mu$ over the
interval $[0,1]$ which in combination with (\ref{eq:5.3}) will
yield the desired upper bound on $\mu$. We use (\ref{eq:5.2}),
(\ref{eq:5.4}) the fact that $p =  \rho$ and $ke^{2\mu} \le 0$ :
\begin{align*}
\int_{0}^{1}\mu(t,r) dr & = \int_{0}^{1}\overset{\circ}{\mu}(r) dr
+ \int_{1}^{t}\int_{0}^{1}\dot{\mu}(s,r) dr ds\\
& \leq C+\int_{1}^{t}\frac{1}{2s}\int_{0}^{1}[e^{2\mu}(8\pi
s^{2}p + k)+1]dr ds\\
& = C+\frac{1}{2}\ln t+ 4\pi\int_{1}^{t}\int_{0}^{1}se^{2\mu}\rho
dr ds
+\int_{1}^{t}\int_{0}^{1}\frac{ke^{2\mu} }{2s} dr ds\\
& \leq C+\frac{1}{2}\ln t+
4\pi\int_{1}^{t}\int_{0}^{1}se^{\mu-\lambda}e^{\mu+\lambda}\rho dr
ds \\
 & \leq C + \frac{1}{2}\ln t + C\int_{1}^{t}\int_{0}^{1}s^3 dr ds\\
& = C + \frac{1}{2}\ln t + Ct^4
\end{align*}
 with (\ref{eq:5.3}) this implies
\begin{equation}\label{eq:5.5}
\mu(t,r) \leq C(1+  \ln t + t^{4}+t^2) \leq Ct^4, \ t \in [1, T[,
\ r \in [0,1].
\end{equation}
And we conclude by theorem \ref{T:4.8} that $T = \infty$.$\square$
\begin{remark}\label{R:5.2}
we have proven that for initial data as in theorem \ref{T:4.4},
the corresponding solution exists for all $t \in [1, \infty[$ and
satisfies the estimates
(\ref{eq:5.2})-(\ref{eq:5.4})-(\ref{eq:5.5}).
\end{remark}

In the case $k=0$ the wave equation can be reduced to a simple linear
equation and the result follows from \cite{rendall95}. Thus it is
the case $k=-1$ of this theorem which is new. The reduction in the
case $k=0$ goes as follows. In that case the field equations imply
that $\lambda-\mu+\log t$ is constant in time. It may, however,
be dependent on $r$. Suppose that $r$ is replaced by a new
coordinate $s$ on the initial hypersurface. Choosing $s$
appropriately makes the transformed quantity  $\lambda-\mu+\log t$
constant on the initial hypersurface and hence everywhere.
Once this transformation has been carried out the wave equation
simplifies to $\ddot\phi+t^{-1}\dot\phi=\phi''$.

     \textbf{\textit{Acknowledgements}} : The authors acknowledge
support by a research grant from the VolkswagenStiftung, Federal
Republic of Germany.

\end{document}